\documentclass[twocolumn,showpacs,preprintnumbers,amsmath,amssymb]{revtex4}

\usepackage{graphicx}
\usepackage{dcolumn}
\usepackage{latexsym}
\usepackage{bm}

\begin{document}

\title{Lattice calculation of low energy constants with Ginsparg-Wilson 
type fermions}
\author{Christof Gattringer}
\email{christof.gattringer@uni-graz.at}
\author{Philipp Huber}
\email{philipp.huber@uni-graz.at}
\author{C.\,B.\ Lang }
\email{christian.lang@uni-graz.at}
\affiliation{{\rm(for the Bern-Graz-Regensburg (BGR)
collaboration)}\vspace{1mm}\\ Institut f\"ur Physik, FB Theoretische Physik\\
Universit\"at Graz, A-8010 Graz, Austria 
}
\date{October 26, 2005}

\begin{abstract}
We present a quenched lattice calculation of low energy constants using the
chirally improved Dirac operator. Several lattice sizes at different lattice
spacings are studied. We systematically compare various  methods for computing
these quantities, using pseudoscalar and axial vector correlators. We find
consistent results for the different approaches, giving rise to $f_\pi =
96(2)(4)\;\textrm{MeV}$, $f_K = 106(1)(8)\;\textrm{MeV}$, $f_K/f_\pi=1.11(1)(2)$,
 $\Sigma= -(286(4)(31)\;
\textrm{MeV})^3$, the average light quark mass  $\bar m =
4.1(2.4)\;\textrm{MeV}$ and  $m_s = 101(8)\;\textrm{MeV}$.
\end{abstract}

\maketitle


\section{Introduction and Motivation}

Low energy theorems were derived quite early in the study of hadrons. When QCD
was accepted as a prime candidate for the   underlying quantum field theory it
became clear that its (approximate) chiral flavor symmetry is spontaneously
broken. One may  use the corresponding ground state as a starting point for a
systematic expansion (Chiral Perturbation Theory, ChPT) taking into account the
explicit symmetry violations due to small masses of the light quarks. In such an
expansion phenomenological parameters, the so-called low energy constants, have
to be determined independently.

The leading order low energy parameters like the pion decay constant or the 
quark condensate are well-known from ``classical'' low energy theorems and
experiments. It is a challenge, however, to find these parameters based 
exclusively on ab-initio calculations for QCD. Of course also QCD has its
minimal set of input parameters (fixing the scale and the quark masses), but
except for these, all other properties can, at least in principle, be derived.
Due to the special nature as a strongly interacting quantum field theory only
non-perturbative calculations lead to that numbers. The lattice formulation
appears to be the prime approach for that aim, using a non-perturbative computer
evaluation of the QCD path integral.

For a long time the lattice approach has been plagued by the problem of
incorporating chiral symmetry in the fermionic action. It took almost two 
decades from the first numerical results for lattice gauge theory before it was
understood how chiral symmetry may be formulated  on the lattice. The
Ginsparg-Wilson condition (GWC) \cite{GiWi82}, characterizes the class of Dirac
operators  allowing for the lattice analog of chiral symmetry \cite{Lu98}. The 
Dirac operators obeying the GWC exactly \cite{NaNe93a,NaNe95,Ne98,Ne98a} or 
approximately \cite{Ka92,FuSh95,HaNi94,Ga01,GaHiLa00} allow one to approach
smaller  quark masses than with the more traditional Wilson fermions. 

However, the GW-fermions need additional effort for their numerical 
implementation. Overlap fermions, which obey the GWC exactly,  are almost two
orders of magnitude more expensive in terms of computer power than Wilson
fermions. The approximate GW-fermions are cheaper, but still one order of
magnitude more expensive. For that reason there are only few first attempts to
use these Dirac operators for a simulation of full QCD with dynamical quarks.
There are, however, several results for the quenched case.

In \cite{GaGoHa03a} hadron masses in the quenched case have been studied for two
types of Dirac operators that obey the Ginsparg-Wilson relation to a good
approximation: the  fixed point operator and the chirally improved operator. Due
to their good chiral behavior, these fermions provide a suitable framework for
the calculation of low energy constants. Since such a calculation involves
quantities that are
renormalized, one also has to determine renormalization constants for the
connection with a continuum scheme like $\overline{\textrm{MS}}$. Whereas  in
the case of the overlap action there are exact symmetries relating the
renormalization constants of the scalar with the pseudoscalar and the vector
with the axial vector sectors, for fermions with only approximate GW-symmetry
such relations have to be checked, too.

For the overlap action there have been several determinations of low energy
parameters
\cite{GiHoRe01E,GiHoRe02,BaBeGa05,GiLuWe03,ChHs03,GiHeLa0304,DoDrHo02,DrDoHo03a,HeJaLe99,HaHaJo02,%
WeWi05,BiSh05,DuHo05}.  
For the chirally improved Dirac operator \ensuremath{D_\textrm{\scriptsize CI}}\
only preliminary results have  been published \cite{GaGoHa03b,GaGoHa03a}. The
reason was, that the renormalization constants were not available. Meanwhile the
necessary constants for quark bilinears  have been determined for
\ensuremath{D_\textrm{\scriptsize CI}}\ in \cite{GaGoHu04}. This now allows us
to compute some of the basic low energy parameters in the quenched case. For
this  purpose we study quenched QCD at various values of the quark masses and
determine results for $m_u=m_d\neq m_s$ in the $u \overline{d} $ and $u
\overline{s} $ meson sector for several lattice sizes and lattice spacings, down
to a pion mass of 330 \textrm{MeV}. Our gauge action is the L\"uscher-Weisz
action \cite{LuWe85}.

In Sect.\;\ref{sec:notation} we fix our notation and recapitulate the basic
relations like PCAC, the axial Ward identity and the Gell'Mann-Oakes-Renner
(GMOR) relation. We then discuss the  observables used in our analysis. In
Sect.\;\ref{sec:technicalities} we discuss the details of the lattice simulation
and present the results  for the masses and low energy parameters (like the
meson and quark masses, meson decay constants, chiral condensate) in 
Sect.\;\ref{sec:results}. We summarize  and conclude in
Sect.\;\ref{sec:conclusion}.

\section{Low energy constants and their relations} \label{sec:notation}

All of the relations are given for Euclidean space-time. We first introduce our
notation and then briefly summarize the three key relations which we need in our
extraction of low energy constants from lattice  simulations. A collection of
the relevant vacuum expectation values and ratios thereof ends this section.

\subsection{Basic definitions and relations}

For flavor symmetry group SU(2) we define the unrenormalized (lattice) 
isovector operators for the pseudoscalar, vector and axial vector sector through
(all fields are  taken at the same space-time point)
\begin{eqnarray}
P^a     &=&\frac{1}{2}\, \overline{\psi} \,\gamma_5 \,\tau^a \,\psi    \;,\\
V_\mu^a &=&\frac{1}{2}\, \overline{\psi} \,\gamma_\mu \,\tau^a \,\psi    \;,\\
A_\mu^a &=&\frac{1}{2}\, \overline{\psi} \,\gamma_\mu \,\gamma_5 \,\tau^a\psi \;.
\end{eqnarray}
The fermion fields are flavor doublets $\psi=(u,\,d)$ and $\tau^a, a = 1,2,3$
are the Pauli-matrices. Charged quark bilinear operators are defined through,
e.g.,
\begin{equation} 
P^+ = \frac{1}{2}\,\overline{\psi}\, \gamma_5 \,(\tau^1-\mathrm{i}\,\tau^2) \,\psi
= \overline{d}  \,\gamma_5\, u \;.
\label{chargedbil}
\end{equation}

Quantities renormalized according to a continuum renormalization scheme are
denoted with a  superscript $(r)$. As our reference scheme we use the
$\overline{\textrm{MS}}$-scheme  at a scale of $\mu= 2$ \textrm{GeV}. The
renormalized quantities are related to their  lattice counterparts via
renormalization constants,
\begin{eqnarray}
P^{(r)a}    &=& Z_P \, P^{a}   \;, \label{zdef1} \\
V_\mu^{(r)a}  &=& Z_V \, V_\mu^{a} \;, \label{zdef2} \\
A_\mu^{(r)a}  &=& Z_A \, A_\mu^{a} \;, \label{zdef3} \\
m^{(r)}     &=& Z_m \, m       \;, \label{zdef4} \\
\Sigma^{(r)}     &=& Z_S \, \Sigma \;\;\textrm{with}\;\;Z_S=1/Z_m\;. \label{zdef5}
\end{eqnarray}
Here $m^{(r)}$ denotes the renormalized quark mass and $\Sigma^{(r)}$ the
renormalized condensate, where the bare condensate reads
\begin{equation}
\Sigma\equiv\frac{1}{2}\langle \overline{u}  u+ \overline{d}  d\rangle\;.
\end{equation}

The renormalization constants relate different schemes.  For exactly chirally
symmetric actions we would have $Z_S=Z_P$ and $Z_V=Z_A$. The chirally improved
Dirac action used here is only approximately chirally symmetric and we have
computed the renormalization factors (in the chiral limit) in \cite{GaGoHu04},
utilizing the non-perturbative methods suggested in \cite{MaPiSa95,GoHoOe99}. 

We now discuss the basic relations in terms of renormalized quantities 
(measurable experimentally and defined in the $\overline{\textrm{MS}}$-scheme).
The axial vector current is related to the vector current by commutation
relations in current algebra. For  conserved vector currents the normalization
is therefore fixed for both. The axial vector operator couples to the weak
interaction currents. Its relation to the physical (renormalized) isovector pion
field defines the pion decay constant, i.e.,
\begin{equation} \label{pcac1}
\partial_\mu A^{(r)a}_\mu=M_\pi^2\,f_\pi\,\phi^{(r)a}\;.
\end{equation}
(There are also other conventions differing by, e.g., a factor of  $\sqrt{2}$.
Our definition corresponds to an experimental value of 92.4(3) MeV
\cite{PDBook04}.) The (renormalized) pion field obeys ($x=(\vec{x},\,t)$)
\begin{equation}\label{pionfieldnormalization}
\langle 0 |\, \phi(x)^{(r)a}\,| {\pi^b(\vec{p}=0)} \rangle
= \delta_{ab}\,\mathrm{e}^{-M_\pi\,t}\;.
\end{equation}

Various equivalent expressions for expectation values may be derived, in
particular the so-called partially conserved axial vector current (PCAC)
relation:
\begin{equation}\label{pcac2}
\partial_\mu\langle 0 |\, A^{(r)a}_\mu(x)\,| {\pi^b(\vec{p}=0)}  \rangle
=
\delta_{ab}\,M_\pi^2\,f_\pi\,\mathrm{e}^{-M_\pi\,t}\;.
\end{equation}

Global symmetries of an action at the classical level lead to conserved Noether
currents. In a quantum field theory global symmetries of the action are  expected
to manifest themselves by an analog to the conservation of Noether currents. To
derive these relations one considers infinitesimal symmetry transformations of the
fermion fields in the path integral
\begin{equation}
     \langle 0|{\cal O}|0 \rangle=\frac{1}{Z}\int dU\, d\bar\psi\,
     d\psi\,
     \;{\cal O}[U, \bar\psi, \psi ]\; e^{-S[U, \bar\psi, \psi ]}\;,
\end{equation}
where $ {\cal O}$ denotes an arbitrary operator. Some classical symmetries may be
spontaneously broken or broken due to anomalies resulting from the functional
integration. If the integration measure is invariant under  the transformation this
lead to (Ward-) identities of the form
\begin{equation}\label{WI1}
\langle 0|\delta {\cal O}|0\rangle-\langle 0|{\cal O}\delta
S|0\rangle=0\;.
\end{equation}
Thus, the quantum analog to the classical conservation law has similar (or equal,
$ {\cal O}=1$) form, but is an operator identity.

For the QCD Lagrangian (renormalized, e.g., in the $\overline{\textrm{MS}}$-scheme)
one exploits the invariance properties under symmetry transformations to derive
Ward identities. Although obtained first on the classical level and on-shell, the
relations hold under quantization and one ends up with local operator identities
for the full quantized theory. (For the singlet axial vector there is an additional
anomaly contribution from the integration measure.)

In the quark sector chiral symmetry is broken explicitly  by the quark mass
matrix ${\cal M}$ and one finds
\begin{eqnarray}
\partial_\mu V^{(r)a}_\mu 
&=& \frac{1}{2} \,\left(\overline{\psi} [\tau^a ,{\cal M}]\psi\right)^{(r)} \;,\\
\partial_\mu A^{(r)a}_\mu 
&=& \frac{1}{2} \,\left(\overline{\psi} \gamma_5 \{\tau^a,{\cal M}\} \psi\right)^{(r)} \;.
\end{eqnarray}
If the quark masses are degenerate, the vector current is conserved and the axial
vector current obeys the axial Ward identity (AWI),
\begin{equation}
\partial_\mu A^{(r)a}_\mu =  2 \, m^{(r)}\,P^{(r)a} \;.
\end{equation} 
Combining this with relation (\ref{pcac1}) gives
\begin{equation} \label{combinedPCACAWI}
2\, m^{(r)}\,P^{(r)a} =f_\pi\,M_\pi^2 \phi^a\;.
\end{equation}

The correlation function of the normalized pion field 
(cf.\ (\ref{pionfieldnormalization})) reads
\begin{equation}
\langle 0 |  \phi^{(r)a}(\vec p =0,t) \phi^{(r)b}(0) | 0 \rangle=\frac{1}{2\,M_\pi}
\,\delta_{ab}\,\mathrm{e}^{-M_\pi\,t}\;.
\end{equation}
The asymptotic behavior of the pseudoscalar field correlator
\begin{equation}
\langle 0 |  P^{(r)a}(\vec p =0,t) P^{(r)b}(0) | 0 \rangle=(G_\pi^{(r)})^2\,\frac{1}{2\,M_\pi}
\,\delta_{ab}\,\mathrm{e}^{-M_\pi\,t}\;,
\end{equation}
is dominated by the pion state as well. We have introduced the relative factor
$G_\pi^{(r)}$, which, from (\ref{combinedPCACAWI}), is
\begin{equation}
|\langle 0|P^{(r)}|\pi(\vec 0)\rangle|=G_\pi^{(r)}=\frac{f_\pi\,M_\pi^2}{2\,m^{(r)}}\;.
\end{equation}

Another Ward identity may be derived taking ${\cal O}=P$, leading to 
\begin{equation}
\Sigma^{(r)}=
-\frac{m^{(r)}}{M_\pi^2} |\langle 0|P^{(r)}|\pi(\vec 0)\rangle|^2\;,
\end{equation}
where we assume degenerate light quark masses for simplicity.  This, combined
with the definition of the pion decay constant  and (\ref{combinedPCACAWI}), leads to
the  usual form of the GMOR relation:
\begin{equation}\label{GMORrelation}
f_\pi^2\,M_\pi^2=-2 \,m^{(r)} \,\Sigma^{(r)} \;.
\end{equation}
The exploitation of the underlying principles has then led to the development of
chiral perturbation theory \cite{We79,GaLe84}. In that context a systematic
expansion of many observables in terms of low energy constants has been derived.
These constants, however, have to be determined either from experiment or from
basic principles, i.e., non-perturbative solution of the underlying field theory
QCD. The lattice formulation of QCD allows such a determination. 

\subsection{Lattice relations}\label{sec:latticerelations}

Let us now express the renormalized quantities through their lattice counterparts,
using the relations (\ref{zdef1}) -- (\ref{zdef5}). Since we need to avoid the
evaluation of disconnected diagrams we restrict ourselves to the isovector
charged bilinears
\begin{equation}\label{chargedbil2}
\frac{1}{2}\,\overline{\psi}\, \Gamma \,(\tau^1-\mathrm{i}\,\tau^2) 
\,\psi =  \overline{d}  \,\Gamma \, u \;,
\end{equation}
where $\Gamma$ is $\gamma_5, \gamma_\mu$ or $\gamma_\mu \gamma_5$. Consequently
from now on we drop the flavor superscript $a$. 

In the simulations we always keep the quark sources (and thus the operator source) 
at a fixed time $t=0$ and compute propagators to all other lattice points. We sum
over the spatial volume of the sink time slice in order to project to zero spatial
momentum $\vec p=0$. 

Taking into account the renormalization factors, the following correlators of
lattice operators and ratios have been studied ($\mu = 4$ refers to the Euclidean 
time direction, the symbol $\sim$ denotes the asymptotic behavior for large $t$):
\begin{eqnarray} 
\label{eq:corrPP}
Z_P^2\,\langle  P(\vec p =0,t) P(0) \rangle
& \sim &\frac{(G_\pi^{(r)})^2}{M_\pi}\,\mathrm{e}^{-M_\pi\,t}\nonumber\\
& = &
\frac{f_\pi^2 M_\pi^3}{4\,(m^{(r)})^2}\,\mathrm{e}^{-M_\pi\,t}\nonumber\\
&=&\frac{M_\pi\, |\Sigma^{(r)}|}{2\,m^{(r)}}\,\mathrm{e}^{-M_\pi\,t}
\;,\\
\label{eq:corrAP}
Z_A\,Z_P\,\langle\, A_4(\vec p=\vec 0,t)\,P(0)\rangle 
&\sim &
G_\pi^{(r)}\,f_\pi\, \mathrm{e}^{-M_\pi\,t}\nonumber\\
& = &
\frac{f_\pi^2\,M_\pi^2}{2\,m^{(r)}}\, \mathrm{e}^{-M_\pi\,t}\nonumber\\
& = &
|\Sigma^{(r)}|\, \mathrm{e}^{-M_\pi\,t}\;,
\end{eqnarray}
\begin{eqnarray}
\label{eq:corrDAP}
 Z_A\,Z_P\,\langle\, \partial_t A_4(\vec p=\vec 0,t)\,P(0)\rangle 
&\sim &
G_\pi^{(r)}\,f_\pi\,M_\pi\, \mathrm{e}^{-M_\pi\,t}\nonumber\\
& = &
\,\frac{f_\pi^2\,M_\pi^3}{2\,m^{(r)}}\, \mathrm{e}^{-M_\pi\,t}\nonumber\\
& = &
|\Sigma^{(r)}|\, M_\pi\, \mathrm{e}^{-M_\pi\,t}\;,\\
\label{eq:corrAA}
Z_A^2\,\langle\, A_4(\vec p=\vec 0,t)\,A_4(0)\rangle 
&\sim&
M_\pi\,f_\pi^2\,\mathrm{e}^{-M_\pi\,t}\;,\\
\label{eq:corrDAA}
Z_A^2\,\langle\,\partial_t A_4(\vec p=\vec 0,t)\, A_4(0)\rangle 
&\sim&
M_\pi^2\,f_\pi^2\mathrm{e}^{-M_\pi\,t}\;,
\end{eqnarray}
\begin{eqnarray}
\label{eq:corrAAPP}
Z_A\,Z_P\,&&\hspace{-4mm}\sqrt{ \langle\,A_4(\vec p=\vec 0,t)\, A_4(0)\rangle \,\langle\, P(\vec p=\vec 0,t)\,  P(0)\rangle}
\nonumber\\
&&\hspace{29mm}\sim
|\Sigma^{(r)}|\,\mathrm{e}^{-M_\pi\,t}\;.
\end{eqnarray}
The pseudoscalar masses thus may be derived from the exponential  decay and the
other low energy parameters from its coefficient.  We cannot use correlators like
$\langle (\partial_t A_4(t))(\partial_t A_4(0))\rangle$  since the source is fixed
to the time slice $t=0$ and thus we cannot construct the lattice derivative there. 

The asymptotic behavior cancels in the following ratios:
\begin{eqnarray}
\label{eq:ratDAXPX}
\frac{Z_A}{Z_P}\,\frac{\langle \,\partial_t A_4(\vec p=\vec 0,t)\, X(0)\,\rangle}
{\langle\, P(\vec p=\vec 0,t)\,X(0) \,\rangle}
&\sim&\frac{M_\pi^2\,f_\pi}{G_\pi^{(r)}}\nonumber\\
&=&Z_m\,2\,m=2\,m^{(r)}
\;,\;\;
\end{eqnarray}
where $X$ may be $P$ or $A_4$. Further useful ratios are
\begin{eqnarray}
\label{eq:ratAPPP}
\frac{Z_A}{Z_P}\,\frac{\langle \, A_4(\vec p=\vec 0,t)\,  P(0)\,\rangle}
{\langle\,  P(\vec p=\vec 0,t)\, P(0) \,\rangle}
&\sim&
\frac{M_\pi\,f_\pi}{G_\pi^{(r)}}=
Z_m\,\frac{2\,m}{M_\pi}\nonumber\\
&=&
\frac{2\,m^{(r)}}{M_\pi}=
\frac{M_\pi f_\pi^2}{|\Sigma^{(r)}|}
\;,\\
\label{eq:ratAAPP}
\frac{Z_A^2}{Z_P^2}\,\frac{\langle \, A_4(\vec p=\vec 0,t)\, A_4(0)\,\rangle}
{\langle\, P(\vec p=\vec 0,t)\, P(0) \,\rangle}
&\sim&
\left(\frac{\,M_\pi\,f_\pi}{G_\pi^{(r)}}\right)^2\nonumber\\
&=&
\left(\frac{\,M_\pi\,f_\pi^2}{\Sigma^{(r)}}\right)^2
\nonumber\\
&=&\left( \frac{2\, m^{(r)}}{M_\pi}\right)^2\;,\\
\label{eq:ratDAPAA}
\frac{Z_P}{Z_A}\,\frac{\langle \,\partial_t  A_4(\vec p=\vec 0,t)\,  P(0)\,\rangle}
{\langle\,  A_4(\vec p=\vec 0,t)\, A_4(0) \,\rangle}
&\sim&
\frac{G_\pi^{(r)}}{f_\pi}=\frac{M_\pi^2}{2\,m^{(r)}}\;,
\end{eqnarray}
\begin{eqnarray}
\label{eq:ratDAPDAPAAPP}
\frac{\langle \,\partial_t A_4(\vec p=\vec 0,t)\,  P(0)\,\rangle^2}%
{\langle\, A_4(\vec p=\vec 0,t)\,A_4(0) \,\rangle \langle\,P(\vec p=\vec 0,t)\, P(0) \,\rangle}
&\sim&
M_\pi^2\;.\;\;
\end{eqnarray}
Ratios involving the lattice derivative $\partial_t A_4$ depend on the way the
derivative is taken. Details will be discussed in Subsection
\ref{subsec:fits_error_deriv}.

Due to lattice periodicity and the parity properties of the meson propagators, the
exponential term $\exp{(-M\,t)}$ is accompanied by another  term from the
propagation backwards in time, $\exp{(-M\,(T-t))}$. Depending on the correlation
function we therefore observe $\cosh$- (for  $\langle P P\rangle$ and $\langle A_4
A_4 \rangle$ correlators)  or $\sinh$-behavior (for $\langle A_4 P\rangle$ and
$\langle (\partial_t A_4) A_4\rangle$ correlators).

\section{Technicalities}
\label{sec:technicalities}

\subsection{Setup}

In the fermion action we use the chirally improved Dirac operator
\ensuremath{D_\textrm{\scriptsize CI}}\  \cite{Ga01,GaHiLa00}. It is based on a
systematic expansion of the lattice Dirac operator taking into account the whole Clifford
algebra and terms coupling fermions within a certain range of neighbors on the
lattice. The expanded Dirac operator is inserted in the GWC which  then leads to a
set of algebraic equations for the expansion coefficients.  The equations include a
normalization condition, which depends on the lattice spacing and thus on the gauge
action. The solution of the system of algebraic equations gives the coefficients
defining \ensuremath{D_\textrm{\scriptsize CI}}. In recent applications
\cite{GaGoHa03a}, as well as here, we use a  set of 19 independent terms in the
expansion in the action. Taking into account the lattice symmetries this then
corresponds to several  hundred coupling terms at each site of the lattice. Finally, in
the  definition of \ensuremath{D_\textrm{\scriptsize CI}}\  also one step of
HYP-smearing of the gauge configuration is included \cite{HaKn01}. For the gauge
fields we use the L\"{u}scher-Weisz  action \cite{LuWe85}. The lattice spacings
have been determined  using the Sommer parameter \cite{GaHoSc02}. We summarize the
simulation parameters in Table~\ref{tab:simulationparameters}. 

The strange quark mass parameter was set using the kaon mass (chiral extrapolation
with  non degenerate quark masses). This gives a value \cite{Ha05a} of  $a\,m_s
=0.089(2)$ for $\beta=7.90$ and $a\,m_s = 0.061(1)$ for $\beta=8.15$. For  $\beta =
8.15$ we have quark propagators at $a\,m = 0.06$ and we used those for  the strange
quark. For $\beta = 7.90$ we have quark propagators for  $am = 0.08$ and $a\,m =
0.10$. We interpolate the hadron results from those two values to the strange quark
mass at $a\,m = 0.089$.  All results for the $K$-meson are based on quark
propagators with non-degenerate quark masses. In the discussion the corresponding
meson fields, however,  will still be called $P$ and $A$ in order to simplify our
notation.

The necessary renormalization constants for fermionic bilinears were computed  in
\cite{GaGoHu04} and we list them in Table~\ref{tab:renormalization_constants} for
completeness. All $Z$ are in the chiral limit and for the 
$\overline{\mbox{MS}}$-scheme at a scale of 2 \textrm{GeV}. The values at $\beta = 8.15$ were obtained by
interpolation.


\begin{table}[t]
\caption{Parameters of the simulation. Where the strange quark mass is  given (in
parenthesis), we also determined propagators for  strange hadrons. Type denotes the
type of the quark source/sink and \#\,cf.\ the number of configurations entering
the analysis. 
\label{tab:simulationparameters}}
\begin{ruledtabular}
\begin{tabular}{rcccccr}
$ L^3\times T $   & $\beta$ & $a[\textrm{fm}]$ & $a[\textrm{GeV}^{-1}]$ & \#\,cf.\ & Type & $a\,m$ $(a\,m_s)$ \\ \hline
$8^3\times24$     & $7.90$  & $0.148$ & $0.750$       & 200 	 & $p,n$    & $0.02-0.20$	\\
$12^3\times24$    & $7.90$  & $0.148$ & $0.750$       & 100 	 & $p,n$    & $0.02-0.20$	\\
$12^3\times24$    & $8.35$  & $0.102$ & $0.517$       & 100 	 & $p,n$    & $0.02-0.20$	\\
$16^3\times32$    & $7.90$  & $0.148$ & $0.750$       &  99 	 & $p,n$    & $0.02-0.20$	\\
$16^3\times32$    & $8.35$  & $0.102$ & $0.517$       & 100 	 & $p,n$    & $0.02-0.20$	\\
$16^3\times32$    & $8.70$  & $0.078$ & $0.395$       & 100 	 & $p,n$    & $0.02-0.20$	\\ 
$16^3\times32$    & $7.90$  & $0.148$ & $0.750$       & 100 	 & $p,n,w$  & $0.02-0.20$	\\
                  &         &	      &               &     	 &          & ($0.08, 0.10$)    \\
$20^3\times32$    & $8.15$  & $0.119$ & $0.605$       & 100 	 & $p,n,w$  & $0.017-0.16$      \\
                  &         &	      &               &     	 &          &  ($0.06$)       
\end{tabular}\end{ruledtabular}
~\\

\caption{Renormalization constants taken from \cite{GaGoHu04}. The values 
for $\beta=8.15$ have been obtained by interpolation.
\label{tab:renormalization_constants}}
\begin{ruledtabular}
\begin{tabular}{llllll}
$\beta$      & $Z_S$     & $Z_V$     & $Z_T$     & $Z_A$     & $Z_P^\textrm{\scriptsize{Sub}}$ 
\\ \hline
7.90     & 1.1309(9) & 0.9586(2) & 0.9944(3) & 1.0087(4) & 1.0281(5) \\
$8.15^*$ & 1.081(1)  & 0.967(1)  & 1.014(1)  & 1.011(1)  & 1.012(3)  \\ 
8.35     & 1.039(1)  & 0.973(1)  & 1.028(2)  & 1.012(1)  & 0.987(4)  \\ 
8.70     & 0.959(2)  & 0.979(1)  & 1.049(1)  & 1.0095(7) & 0.915(1)  \\ 
\end{tabular}\end{ruledtabular}
\end{table}

\subsection{Sources and normalization}

In order to improve signals we use Jacobi-smeared \cite{Gu89,BeGoHo97} sources at
$t=0$ for the quarks. For most runs we have one smearing width (denoted as
$n(arrow)$, cf.\ Table~\ref{tab:simulationparameters}). For some data sets, that
have also been used for an analysis of the excited hadron masses \cite{BuGaGl04a},
we have two different widths ($n$ and $w(ide)$).  In addition we also use
point-like sinks ($p(oint)$) for sake of normalizing our results as discussed below. The
smearing widths are chosen such that the effective size of a given source is
approximately the same for all lattice spacings. The parameters (smearing steps and
coupling) thus depend on the gauge coupling \cite{BuGaGl04a}.

For computing the coefficients of the exponential decay necessary for determining
the pion decay constant and the condensate, we have to normalize the hadron sources
to point-like bilinears. Our normalization procedure  is set up as follows: Let us
denote by $X_{s_1 s_2}$ a mesonic operator built out of an  anti-quark of smearing
type $s_1$ and a quark of smearing type $s_2$ with $s_i = n, w$ or $p$. Guided by
the arguments discussed below we find that we may safely assume factorization of the
normalization factors,  
\begin{equation} 
X_{s_1 s_2} = C_{s_1}^X \, C_{s_2}^X \, X_{pp} \;.
\label{factorization}
\end{equation}
For ratios like
\begin{equation}
\frac{\langle X_{s_1 s_2} X_{s_3 s_4} \rangle}{\langle X_{s_1 s_2} X_{pp} \rangle} \;,
\end{equation}
we find excellent plateaus for $t,\,t'\in[t_a, t_b]=[4,\,T-4]$ and violations of the factorization
hypothesis (\ref{factorization}) smaller than 1.2 \% for the narrow and wide
results at $\beta=7.90$, and less than 2.2 \% for the results at $\beta=8.15$.  In
this way we obtain numbers for $(C_n^X)^2$, $(C_w^X)^2$ and $C_n^X C_w^X$ and from
them find the values for $C_n^X$, $C_w^X$. The coefficients are calculated for all
operators $X$, and separately for each gauge coupling, volume and quark mass. For
the large physical volumes we find very little  ${\cal O}(1\%)$ dependence of the
$C_s^X$ on the quark mass which increases for the smallest volumes to a variation
of  ${\cal O}(5-10\%)$ over the range of quark masses we consider.  The  relative
statistical errors from the plateau fits are less than 0.0005 and were neglected in
the further analysis.

As a consistency check we compared the final results for all possible smearing 
combinations and checked whether the resulting masses and in particular the
prefactors of the exponential decay are in agreement when normalized appropriately.
This was the case and in the subsequent presentation we therefore do not discuss
this issue any more. All results presented here are based on the results with
narrow quark sources.

\subsection{Fits, error estimates and numerical derivatives}
\label{subsec:fits_error_deriv}

{\noindent \bf Fitting procedure:}
At larger $t$, where excited state contributions become negligible,  the
propagators have a $\cosh$ or $\sinh$ functional behavior. Therefore we may extract
the prefactor $D$ and the meson mass $M$ performing a correlated least-squares fit
of the correlation function $C(t)$ to
\begin{equation} \label{correlatorfunction}
D(M)\,f(M,\,t)\quad\textrm{with}\quad 
f(M,\,t)\equiv \left(\mathrm{e}^{-M \,t}\pm\mathrm{e}^{-M\,(T-t)}\right) \;,
\end{equation}
by minimizing
\begin{eqnarray}
\chi^2 &=& \sum_{t,t'} \Big[ C(t) - D(M)\,f(M,\,t) \Big] \,\mathrm{Cov}^{-1}(t,\,t') \nonumber\\
&&\;\;\;\times\,\Big[C(t') -D(M)\,f(M,\,t') \Big] \;.
\label{chi2_corr_fit} 
\end{eqnarray}
Here $\mathrm{Cov}$ denotes the covariance matrix for the correlation function  entries
and $t,\,t'\in[t_a, t_b]$. The minimization may be simplified by  observing, that
for given $M$ the minimum of $\chi^2$ is obtained for
\begin{equation}
D(M) = \frac{\sum_{t,t'}\; C(t)\, \mathrm{Cov}^{-1}(t,t')\, f(M,\,t')}{\sum_{t,t'}\; f(M,\,t) 
\, \mathrm{Cov}^{-1}(t,t')\, f(M, t')}
\label{z_corr_fit}\;.
\end{equation}
Using this relation one performs the one-dimensional minimization of
(\ref{chi2_corr_fit}) with regard to $M$.

Ratios of correlation functions are fitted to constants in  $t\in[t_a, t_b]$.

{\noindent \bf Error estimation:}
The correlator values are arranged in overlapping blocks according to the standard
jackknife algorithm. Each block consists of typically 95\% of all hadron
propagators for a given set of lattice parameters. 

For each such jackknife block we then determine the values of the propagator and
the requested ratios at various $t$, as well as the  covariance matrix for the
propagator and the variance for the ratios. These are then fitted as discussed,
i.e.,  either to asymptotic $\cosh$- or $\sinh$-behavior or to a constant. For this
fit the relative weights (as defined from the covariance and variance) are
important.

We also need variances for the correlators involving derivatives with regard to $t$
(discussed below), as they also enter some of the ratios. Since we need these for
each jackknife set we estimate the variance of the  derivatives by performing
another jackknife analysis within the given set.

The fits (and the extrapolations to the chiral limit) are then repeated for all
jackknife blocks and the variation of the results for coefficients, mass values and
ratios leads to the estimate for the corresponding errors.

{\noindent \bf Numerical derivatives:}
For some of the ratios of correlators we need derivatives of the correlator  with
regard to time $t$. Numerical derivatives are always based on assumptions on
the interpolating function. Usual simple 2- or 3-point formulas assume polynomials
as interpolating functions. We can do better by utilizing the information on the
expected $\cosh$- and $\sinh$-dependence. In fact, we may use these functions for
local 3-point  interpolation and get the derivative therefrom. 

A 2-point derivative based only on function values $y_t$ and $y_{t-1}$ is not
suitable since it provides values at half-integer $t$. We therefore use a local
3-point fit to $y_{t-1}$, $y_t$ and $y_{t+1}$ to the functional form
(\ref{correlatorfunction}) (depending on whether the correlator is symmetric or
anti-symmetric in $t$),  where the parameters $\overline D$ and $\overline M$ now
depend on the actual value of $t$. We then reconstruct the derivative as
\begin{equation} \label{correlatorfunctionderiv}
\partial_t f(\overline M,\,t)\equiv
 \overline M \left(-\mathrm{e}^{-\overline M \,t}\pm \mathrm{e}^{-\overline M\,(T-t)}\right)\;.
\end{equation}
Then the desired ratios can be computed in a straightforward manner  and the
plateau values obtained by fits.

When analyzing not ratios but the correlators of type $\langle (\partial_t A_4)
X\rangle$ directly, we perform a global fit to $\langle A_4 X\rangle$ according to
(\ref{correlatorfunction}) and take the analytic derivative. 

\section{Results}\label{sec:results}

Throughout the analysis we refer to the quantities as defined in 
Sect.\;\ref{sec:latticerelations}. Where we give physical, renormalized
($\overline{\textrm{MS}}$-scheme) values we always use the renormalization factors
as obtained  in the chiral limit, even in plots for non-vanishing bare quark
masses. The renormalization factors $Z$ relate to the continuum
$\overline{\textrm{MS}}$-scheme at a scale of $\mu= 2\;$GeV and are $a$-dependent
as given in Table~\ref{tab:renormalization_constants}.

\subsection{Meson Masses}

\begin{figure}[tp]
\includegraphics*[width=8.3cm,clip]{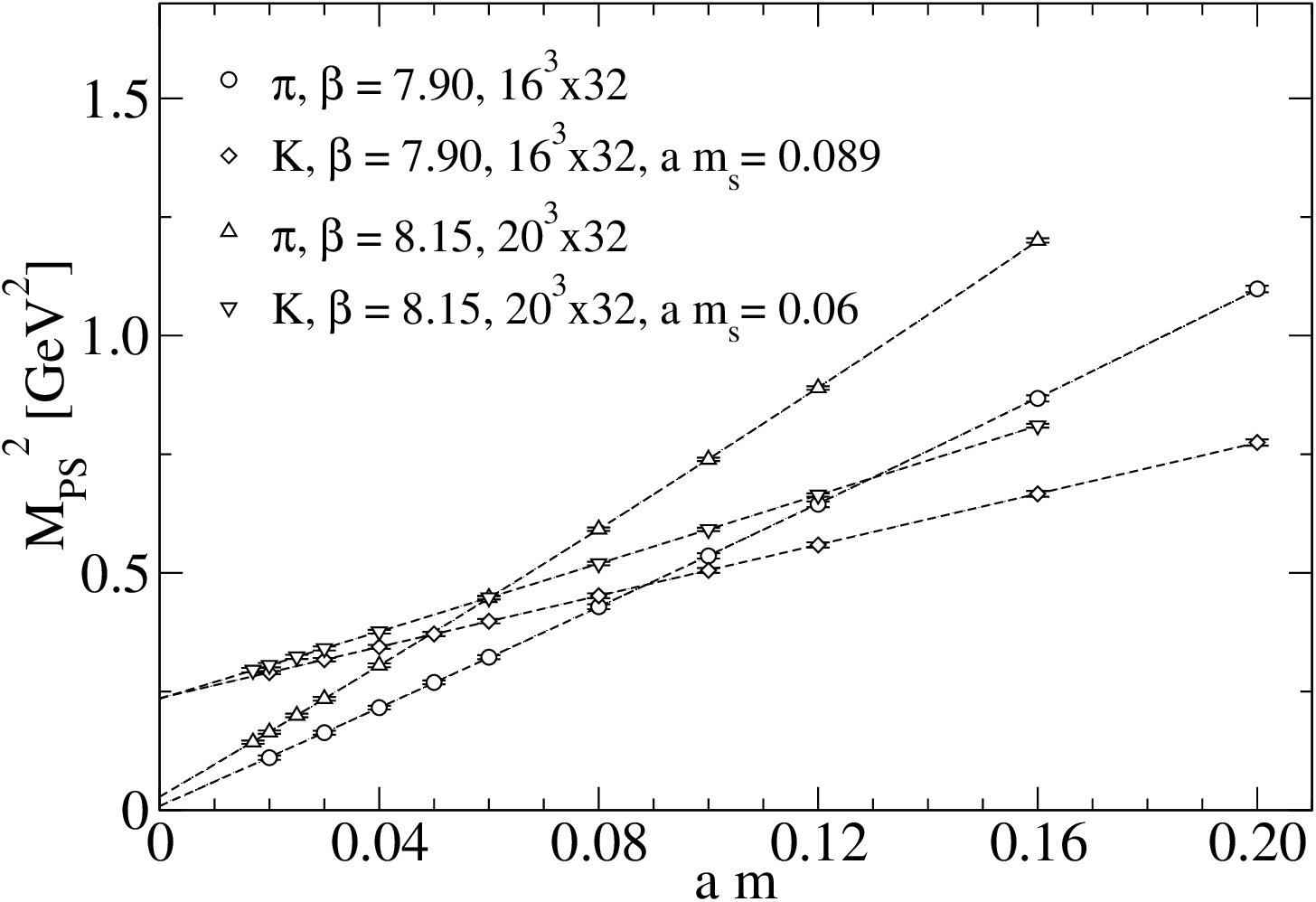}
\caption{{}$M_\pi^2$ and $M_K^2$ determined from the asymptotic behavior of the
$A_4$-correlator (\ref{eq:corrAA}). }
\label{fig:MPS2_vs_am}
\end{figure}

In Fig.\;\ref{fig:MPS2_vs_am} we present results for the pseudoscalar masses as
determined from the exponential fit of the $A_4$-correlator (\ref{eq:corrAA}) for
the time components of the axial vector. The kaon propagators are determined from
the same interpolators but for non-degenerate quark masses. The translation of
results in lattice units to physical mass values was done  using the values for the
lattice constant $a$ given in Table~\ref{tab:simulationparameters}. It is
interesting to note that the linear extrapolation of our pion mass results implies a small
residual quark mass. We remark that when using the pseudoscalar correlator at
$\beta=7.90$ and an extrapolation formula inspired by quenched chiral perturbation
theory   (QChPT) \cite{BeGo92,Sh92,BeGo94}  no such effect was observed
\cite{GaGoHa03a}.

\begin{figure}[tp]
\includegraphics*[width=8.3cm,clip]{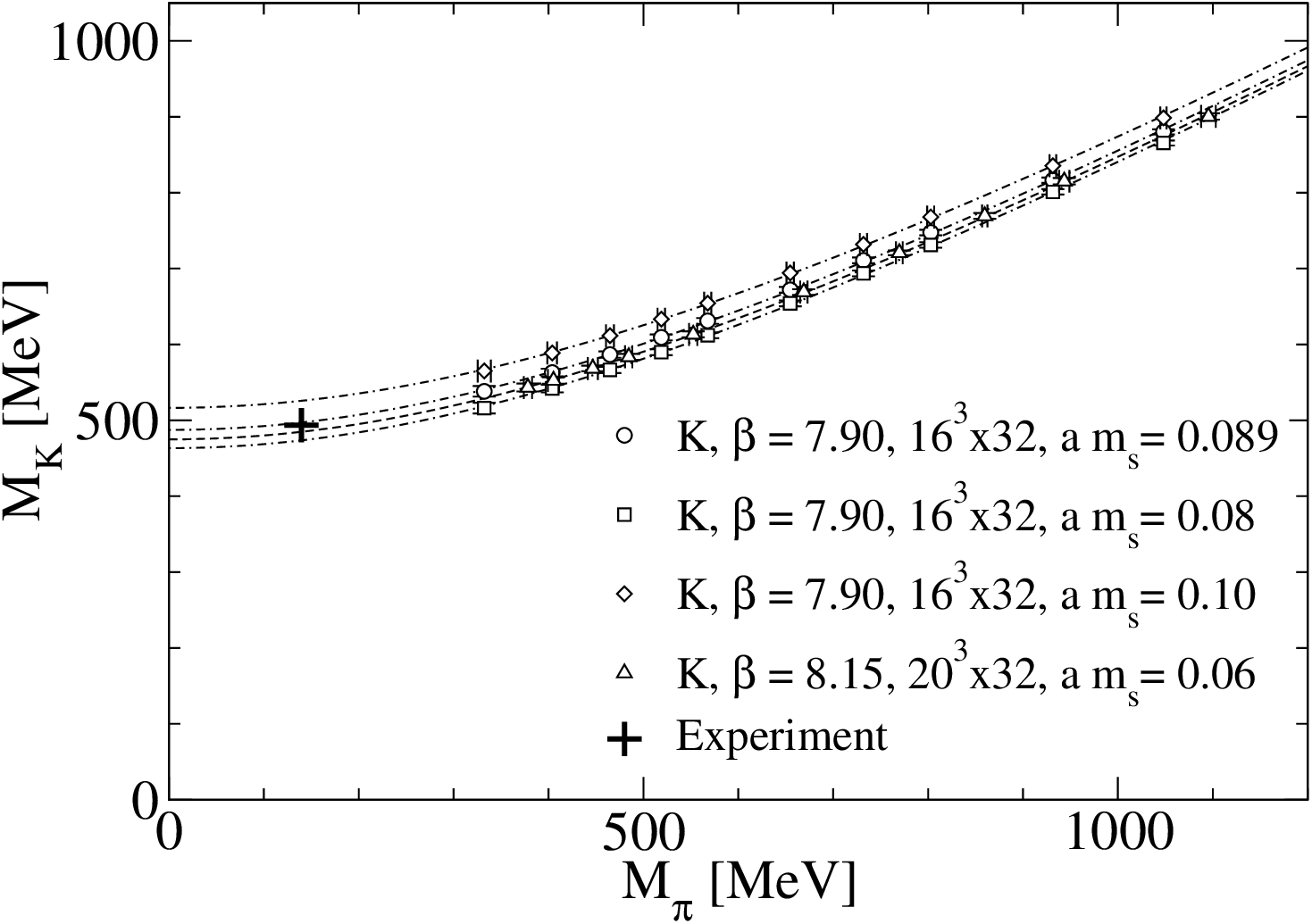}
\caption{{}$M_K$ vs.\ $M_\pi$. 
For the $16^3\times 32$ lattice we have interpolated between two values 
of the strange quark mass; for the $20^3\times 32$ lattice $a\,m_s = 0.06$ 
corresponds already to the physical point.
The curves represent the linear fit in Fig.\;\ref{fig:MPS2_vs_am}.
}
\label{fig:MK_vs_mpi}
\end{figure}

The value of the strange quark mass has been chosen such that the kaon mass agrees
with  experiment at the linear extrapolation in the light quark mass to the
physical point. For the $16^3 \times 32$ lattice at $\beta=7.90$ we interpolate
$M_K^2$ linearly between the two values of the strange quark mass neighboring the
physical point at $a\,m_s = 0.089$, cf.\ Fig.\
\ref{fig:MK_vs_mpi}.

Let us now discuss possible sources of systematic errors.

\begin{figure}[tp]

\includegraphics*[width=8.3cm]{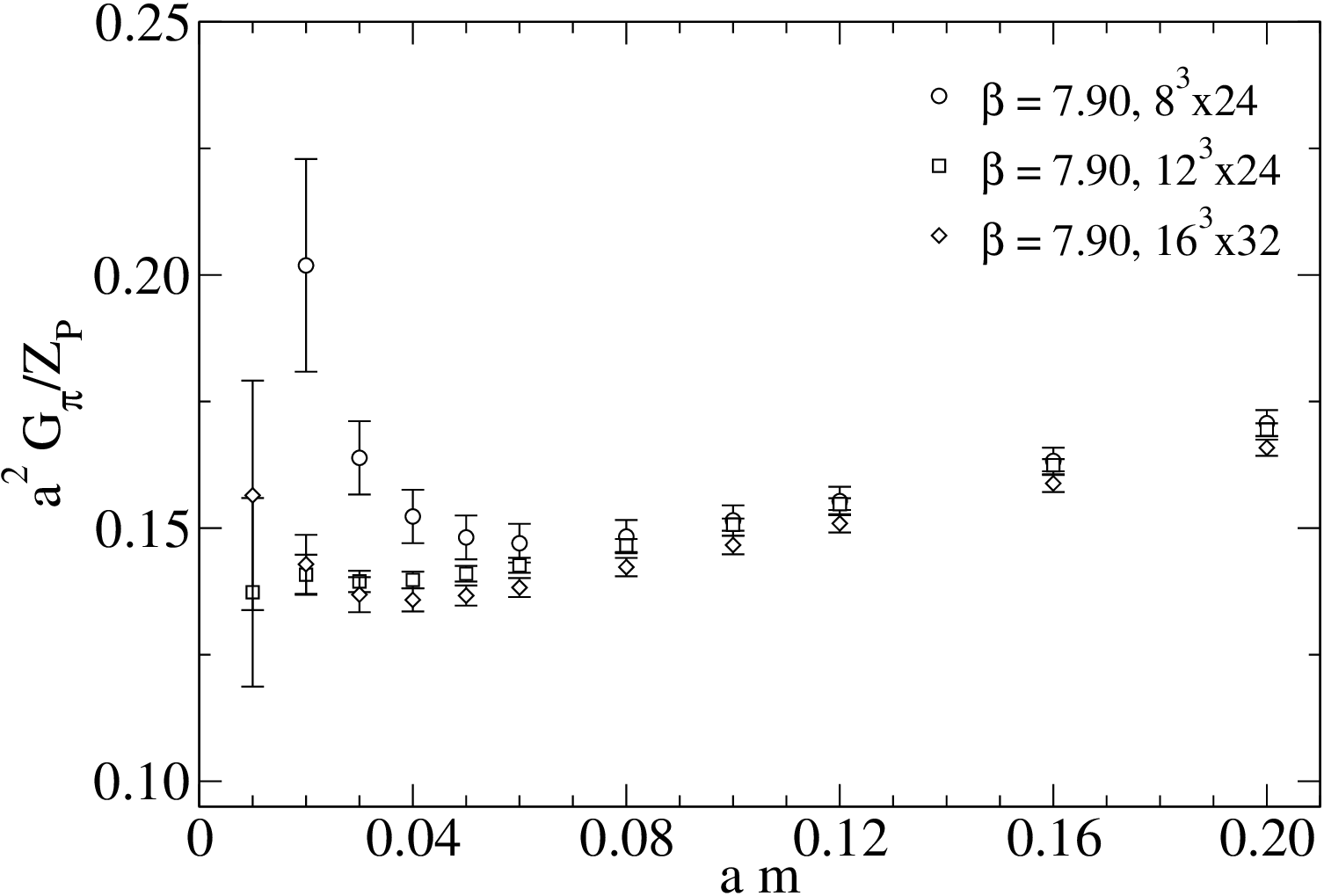}
\caption{\label{fig:Gpi_vol_dependence} $a^2\,G_\pi^{(r)}/Z_P$ from (\ref{eq:corrPP}) for $\beta=7.90$
and lattice sizes $8^3\times 24$, $12^3\times 24$, $16^3\times 32$. The finite
size  significantly affects the small volume data ($8^3\times 24$) for $a\,m
\leq 0.06$. For the other volumes finite size effects become important for $a\,m
< 0.02$.}
\end{figure}

{\noindent \bf Finite size dependence:}
In Fig.\;\ref{fig:Gpi_vol_dependence} we compare $a^2\,G_\pi^{(r)}/Z_P$ determined
from (\ref{eq:corrPP}) for $a\,m\geq 0.01$  for three volumes but at the same
lattice scale. We find a strong volume dependence for the
smallest lattice at bare quark masses below $0.06/a$. This is to be expected since
for that quark masses the inverse pion mass becomes larger than half the spatial
lattice size. For the larger lattices finite volume effects are significant only
for $a\,m<0.02$. We will  therefore discuss only results for the large lattices and
$a\,m\geq 0.02$ (0.017 for the $20^3\times32$ lattice).  The lattices largest in
physical units and of similar spatial extent ${\mathcal O} (2.4\;\textrm{fm})$ are
those of size $16^3\times 32$ at $\beta=7.90\, (a = 0.148\; \textrm{fm})$ and
$20^3\times 32$ at $\beta=8.15\, (a = 0.119\;\textrm{fm})$.

{\noindent \bf Topological finite size effects:}
Exact Ginsparg-Wilson Dirac operators have exact zero modes. In quenched
simulations these are not suppressed by the fermionic determinant. However, their
effects  for, e.g., the pion correlators \cite{Sh97Sh00}, are hard to detect unless
one approaches very small pion masses ${\cal O}(250\;\textrm{MeV})$ 
\cite{DoDrHo04}. For approximate GW-operators like the one studied here the main
problem is the (scarce) occurrence of slightly misplaced zero modes, real
eigenvalues of \ensuremath{D_\textrm{\scriptsize CI}}\,  below zero. These lead to
contributions in the quark propagator that diverge already at some positive, albeit
small, mass. The smaller the quark mass parameter is, the  stronger one might
notice such distortions often referred to as topological finite size effects.

There have been various suggestions to deal with the problem of zero modes in the
pseudoscalar propagator \cite{BlChCh04,GiHoRe01E,DoDrHo02,GaGoHa03a},  among them
the proposal to study a combination of  the iso-non-singlet pseudoscalar and scalar
correlators such that the zero modes cancel. Indeed, zero modes contribute
differently to different propagators. For exactly chiral Dirac operators for the
$A_4$-correlator we expect a contribution of $\mathcal{O}(1/m)$, whereas for the
$P$-correlator we expect $\mathcal{O}(1/m^2)$. 

\begin{figure}[tp]
\includegraphics*[width=8.3cm]{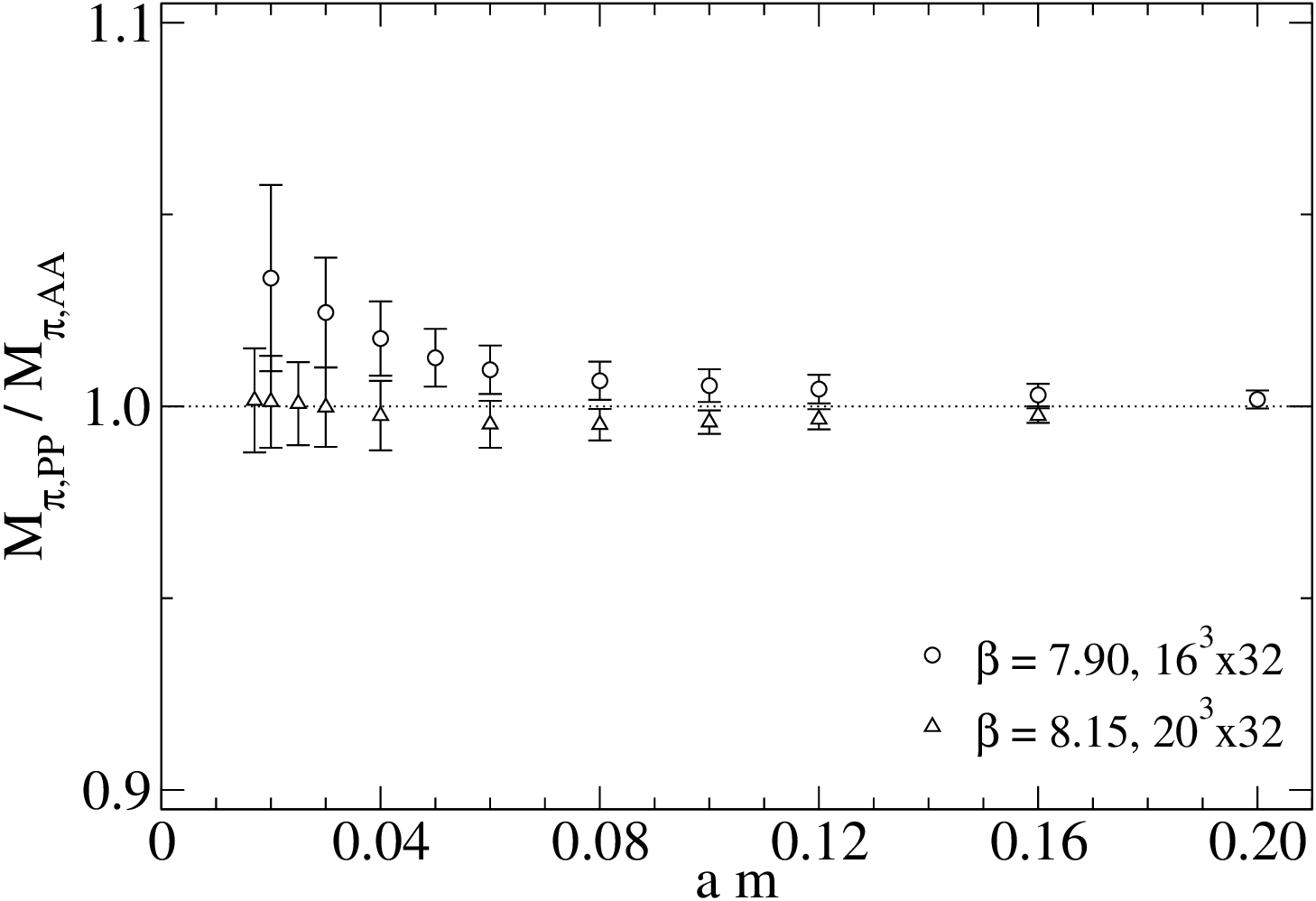}
\caption{\label{fig:mass_ratio} Ratio of the pion masses obtained from the
pseudoscalar propagator $\langle P P \rangle$ and the axial propagator $\langle
A_4 A_4 \rangle$ for the same physical volumes but two different lattice
spacings.}
\end{figure}

In Fig.\;\ref{fig:mass_ratio} we show the ratio of pseudoscalar masses obtained
from the pseudoscalar and axial propagator  ${M_\pi(\langle\, P\, P\,
\rangle)}/{M_\pi(\langle\,A_4\, A_4\,\rangle)}$,  comparing results at two
different lattice spacings for the same physical volumes. There are indications for
zero mode contributions being stronger for  the $P$-correlator on the coarser
lattice. For the final pseudoscalar mass values we use the less affected results
from the $A_4$-correlators.

{\noindent \bf Chiral logs:}
Since quenched QCD does not include the full fermion dynamics, corrections to the
chiral expansion introduce new terms, including singular ones (quenched chiral
logarithms). On the other hand, it has been hard to clearly identify these in
numerical calculations \cite{GaGoHa03a,DoDrHo04,BaBeGa05}. In particular, one needs to
approach parameter regions, where the pion mass is well below 300 \textrm{MeV}. The
situation then is furthermore obscured by the role of zero modes (depending on the
fermion action), poor statistics and variation in the fit range.

In quenched chiral perturbation theory  (QChPT) \cite{BeGo92,Sh92,BeGo94} one
expects 
for the pseudoscalar mass non-analytic behavior in the quark mass parameter
\cite{Sh97Sh00,HeSoWi00}, which arises from hairpin diagrams and
$\eta'$ would-be-Goldstone bosons
\begin{equation}
(a\, M_P)^2 \propto (a \,m)^{1/(1+\delta)} \;,
\end{equation}
where values for $\delta$ ranging from 0.19 to 0.23 have been quoted in recent
literature \cite{GaGoHa03a,DoDrHo04,BaBeGa05}. In the range of values studied here we
cannot disentangle reliably such effects from the possible spurious zero mode
contributions and consequently do not attempt to determine $\delta$.

\subsection{Quark masses}

\begin{figure}[t]
\includegraphics*[width=8.3cm]{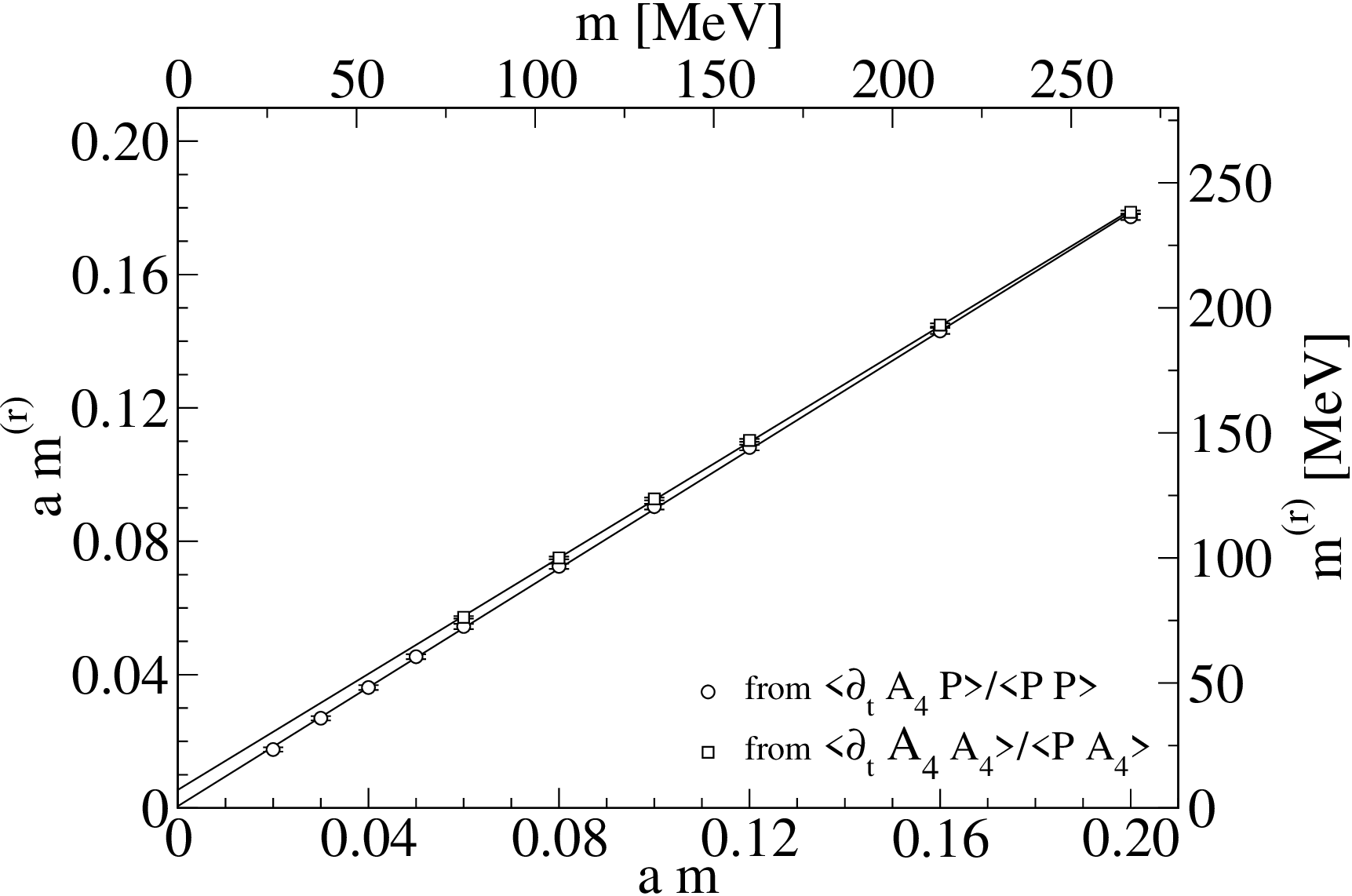}
\caption{\label{fig:amr_vs_am.pion.16x32}
The renormalized quark mass $a\,m^{(r)}$ (in the $\overline{\textrm{MS}}$-scheme)
vs.\ the bare mass parameter $a\,m$ as determined from (\ref{eq:ratDAXPX})
($16^3\times 32$, $\beta=7.90$) using $X = P$ (squares) and $X =
A_4$ (circles). For $X = A_4$ values at quark masses below $a\,m = 0.06$ have
been omitted due to unstable plateau fits.}
\end{figure}

In Eq.\;(\ref{eq:ratDAXPX}) one utilizes the axial Ward identity to obtain the
so-called AWI-mass $m_\textrm{\scriptsize AWI}$; the asymptotic exponential
behavior of numerator and denominator cancels and thus, including the
renormalization factors, this allows one to obtain the renormalized quark mass.
In Fig.\;\ref{fig:amr_vs_am.pion.16x32} we plot the results for two choices of
the pseudoscalar interpolator $P$, showing the results both in lattice and in
physical units.

For the second choice, $X=A_4$, the correlators are of $\sinh$-type and
therefore the ratio becomes numerically unstable near the symmetry point in
$t$. We thus cannot reliably (with small errors) determine the plateau values
for this choice if the quark mass parameter is small. However, at all masses
where we can compare the two results they are in excellent agreement as may be
seen in the figure. The subsequent analysis is based on the more stable choice
$X=P$.

\begin{figure}[t]

\includegraphics*[width=8.3cm]{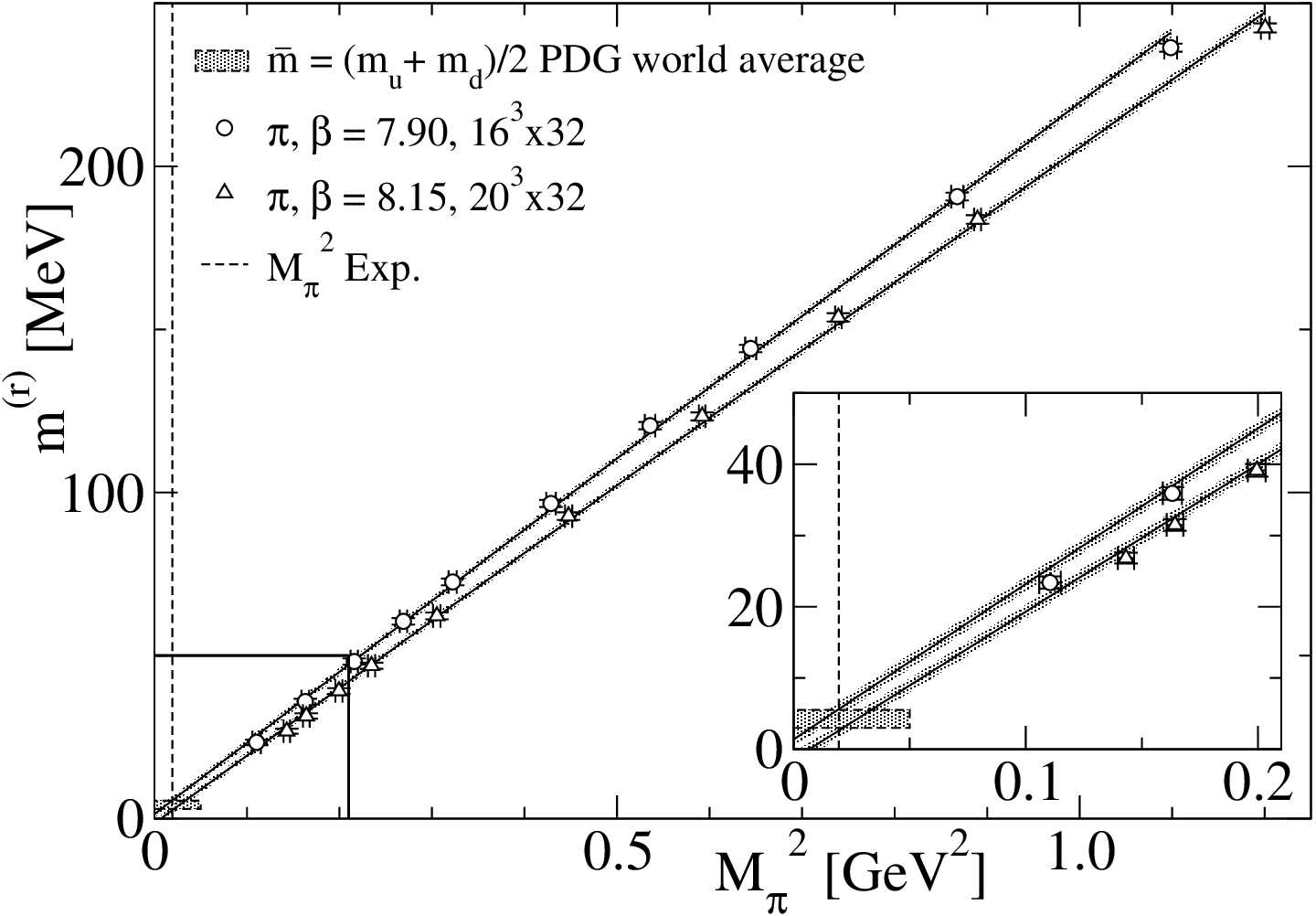}
\caption{\label{fig:mr_vs_Mpi2.pion}
Plot of $m^{(r)}$ (obtained from (\ref{eq:ratDAXPX})) vs.\ $ M_\pi^2$ for two different
lattice spacings but similar physical spatial lattice volume. The linear 
extrapolations are in good agreement with the expected light current quark masses.}
\end{figure}

In Fig.\;\ref{fig:mr_vs_Mpi2.pion} we compare the values, again obtained
utilizing (\ref{eq:ratDAXPX}) with $X=P$. The are derived for two different
lattice spacings and given in physical units. The lines correspond to linear
fits  $m^{(r)}\propto M_\pi^2$, enforcing the simultaneous chiral limit of both
observables. The data do not show deviation from that behavior, although
logarithmic corrections are expected due to quenching. We observe that the
linear extrapolations are in good agreement with the Particle Data Group
\cite{PDBook04}  average for the light quark masses at the physical pion mass.
Averaging the extrapolated values obtained  at the physical pion mass we obtain
an average light quark mass of
\begin{equation}\label{eq:mlight}
\frac{1}{2}\left(m_u^{(r)}+m_d^{(r)}\right)\equiv\bar m^{(r)}
\simeq 4.1(2.4)\;\textrm{MeV}
\end{equation}
in the $\overline{\textrm{MS}}$-scheme.
The error is essentially due to the residual quark mass (see Fig.s\;\ref{fig:mr_vs_Mpi2.pion}
and \ref{fig:MPS2_vs_am}). The precise determination of this residual quark mass is
obscured by the possible contribution of a quenched chiral log 
(cf.\ the discussion in \cite{GaGoHa03a}). This effect is stronger for the smaller physical
volumes and thus we refrain from a determination of $\bar m^{(r)}$ for the
lattices at $\beta=8.35$ and $8.7$. This prohibits a continuum extrapolation 
linear in $a$ and we therefore quote the average of the two values for the two lattices
with spatial extension 2.4 fm.

\begin{figure}[tp]

\includegraphics*[width=8.3cm]{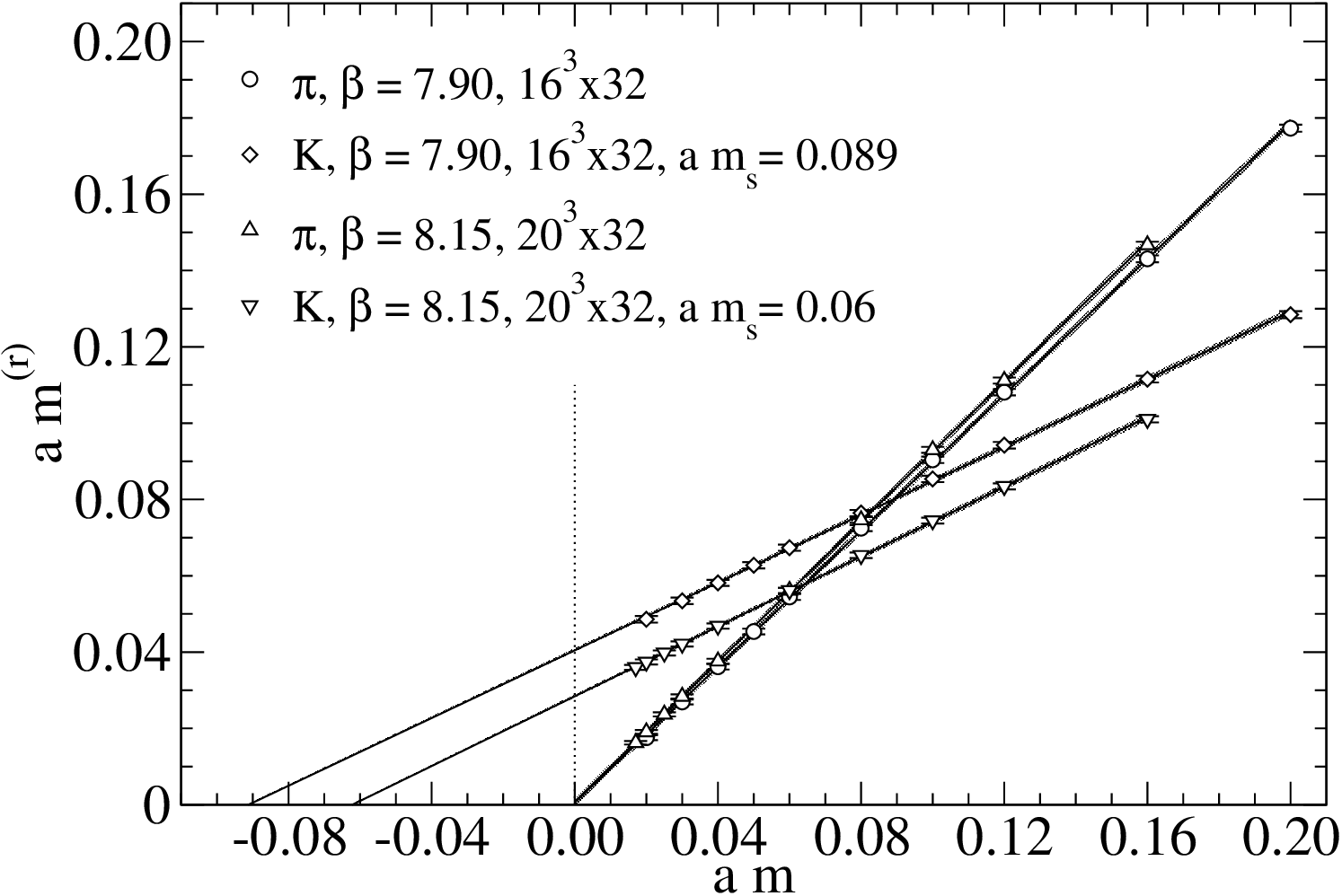}
\caption{\label{fig:amr_vs_am}
$a\,m^{(r)}$ vs.\ $a\,m$ from (\ref{eq:ratDAXPX}) using $X = P$.
The slope of $m^{(r)}$ for the pion data provides $Z_m$.}
\end{figure}

As addressed in Sect.\;\ref{sec:technicalities}, the bare mass parameter for the
strange quark has been fixed such that the kaon mass assumes its physical value
when the data are extrapolated to the physical pion mass \cite{Ha05a}. 
Fig.\;\ref{fig:amr_vs_am} compares the results for the corresponding renormalized
masses (in lattice units) derived from  (\ref{eq:ratDAXPX}) for kaon and pion
correlation functions. For the light quark masses (the pion propagator) the results
confirm those of Fig.\;\ref{fig:amr_vs_am.pion.16x32}, now for two different
lattice spacings superimposed. For the strange quark the fitted lines are close to
being parallel with negative intercepts, in good agreement with the negative bare
strange quark mass parameters.  This again demonstrates that our Dirac operator has
very small additive  mass renormalization.

For the pion data, the slope of $m^{(r)}$ provides a value for the light quark mass
renormalization factor $Z_m$. In Table~\ref{tab:Zmcomparison} we compare these 
numbers with the values of $1/Z_S$ from \cite{GaGoHu04} and find very good
agreement.  From the PCAC-relation and (\ref{eq:ratDAXPX}) we expect for the kaon
data $a\,m^{(r)}\simeq Z_m\,a\,(m_s+m)/2$ which is indeed observed.

\begin{table}[bt]

\caption{Renormalization constant $Z_S$ from \cite{GaGoHu04}  (see Table
\ref{tab:renormalization_constants}) compared to the values of $Z_m$ as derived
from the slope of the renormalized quark mass.}
\label{tab:Zmcomparison}
\begin{ruledtabular}
\begin{tabular}{lllllll}
  $\beta$ & $a [\textrm{fm}]$ & $a [\textrm{GeV}^{-1}]$ & $Z_S$ &     $1/Z_S$   & $Z_m$    & $Z_m Z_S$ \\
\hline
  7.90 &    0.148 & 	0.750 & 	  1.1309(9) & 0.8842(7) & 0.891(4)  & 1.007(5) \\ 
  8.15 &    0.119 & 	0.605 & 	  1.081(1)  & 0.9250(9) &  0.916(5) & 0.991(6) \\ 
\end{tabular}\end{ruledtabular}
\end{table}

\begin{figure}[tp]
\includegraphics*[width=8.3cm]{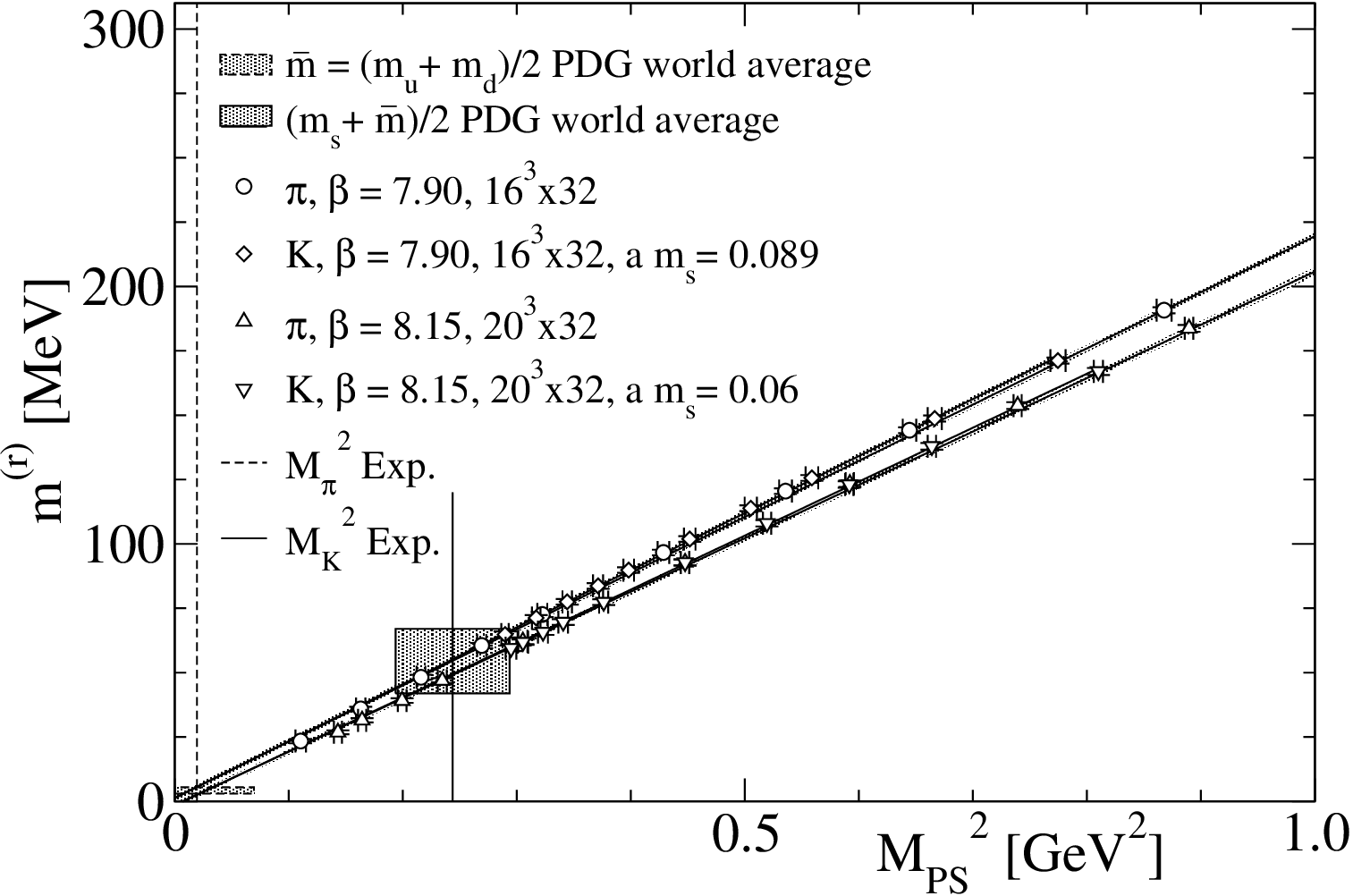}
\caption{\label{fig:mr_vs_MPS2}
$m^{(r)}$ vs.\ $M_{PS}^2$ from (\ref{eq:ratDAXPX}) using $X = P$. Here $M_{PS}$ denotes $M_\pi$ or
$M_K$, respectively. All masses are given in physical units in the $\overline{\textrm{MS}}$-scheme.}
\end{figure}

The quark mass data shown in Fig.\;\ref{fig:mr_vs_MPS2}, now in physical units in the
$\overline{\textrm{MS}}$-scheme, again give a very consistent picture. Note that
now the abscissa gives the corresponding pseudoscalar mass, i.e., that of the pion
or the kaon, for the corresponding values of $m^{(r)}$. The error bands show that
for given lattice spacing the numbers for kaon and pion are on top of each other,
although these states have different quark content. (This justifies the often used
strategy to obtain kaon results from pion  propagators using degenerate mass
quarks.)

Like for the light quark masses in Fig.\;\ref{fig:mr_vs_Mpi2.pion}, also the result for the renormalized
strange quark mass is in excellent agreement with the world average 
\cite{PDBook04}. Averaging the two values computed at the physical kaon mass for
the two lattice spacings we obtain for the strange quark mass 
\begin{equation}
\frac{1}{2}\left(m_s^{(r)}+\bar m^{(r)}\right)\simeq 52(3)\;\textrm{MeV}
\end{equation}
in the $\overline{\textrm{MS}}$-scheme. With (\ref{eq:mlight}) for the light quarks
this then gives $m_s^{(r)}=101(8)~\textrm{MeV}$. Possible finite size effects and
other systematic effects  like chiral extrapolation and quenching have not been
accounted for. The given error takes into  account the standard error and includes
the derivations due to the  dependence on the lattice spacing. 
As discussed above for the light quarks, we have only values at two lattice spacings 
available and therefore cannot perform a sensible continuum extrapolation. 
Our numbers for the average are in
good agreement with determinations from the overlap action in 
\cite{GiHoRe02,ChHs03}, and to a less extent also with \cite{DuHo05}.

\subsection{Condensate}

For exact GW-fermions the chiral condensate may be obtained from the
trace of the inverse Dirac operator. In our case the subtraction constant is not
known to sufficient precision and therefore this approach, successfully applied for
the overlap operator \cite{HeJaLe99,HaHaJo02}, does not work \cite{Ho04a}. Another
method uses the density distribution of the small complex eigenvalues; this would
be a promising method in our case but it requires the costly determination of the low lying 
eigenvalue spectrum for all configurations which we did not calculate.

\begin{figure}[t]
\includegraphics*[width=8.3cm]{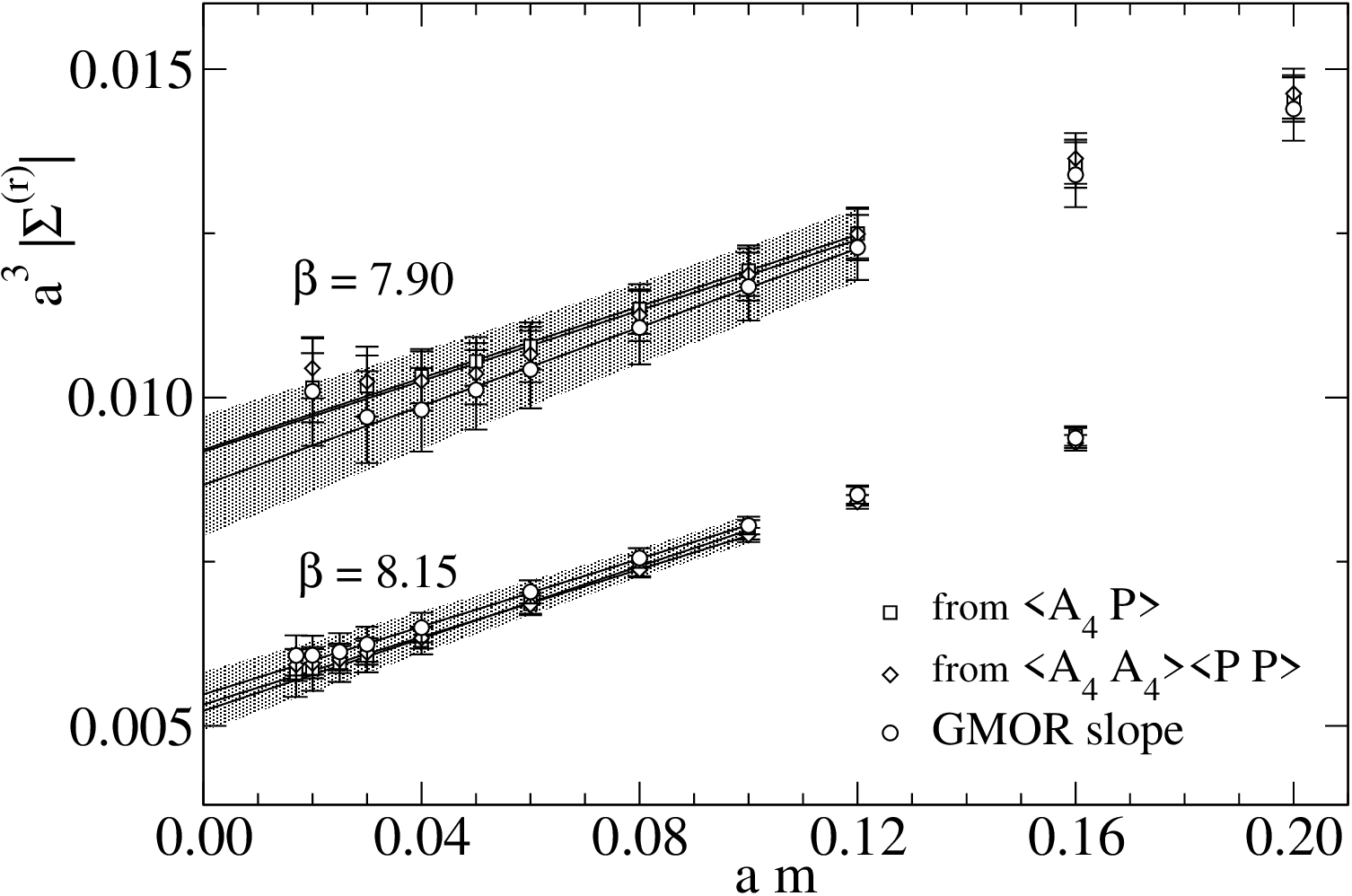}
\caption{\label{fig:a3sigma_vs_am}
$a^3\,|\Sigma^{(r)}|$ vs.\ $a\,m$ for $16^3\times 32$, $\beta=7.90$ and 
$20^3\times 32$, $\beta=8.15$ comparing different 
way of extraction; 
circles: GMOR-slope Eq.\;(\ref{GMORrelation}),
boxes: $\langle A_4 P\rangle$ Eq.\;(\ref{eq:corrAP}),
diamonds: $\langle A_4 A_4\rangle\langle P P\rangle$ Eq.\ (\ref{eq:corrAAPP}).
We also show linear extrapolations to the chiral limit.
}
~\\  
\includegraphics*[width=8.3cm]{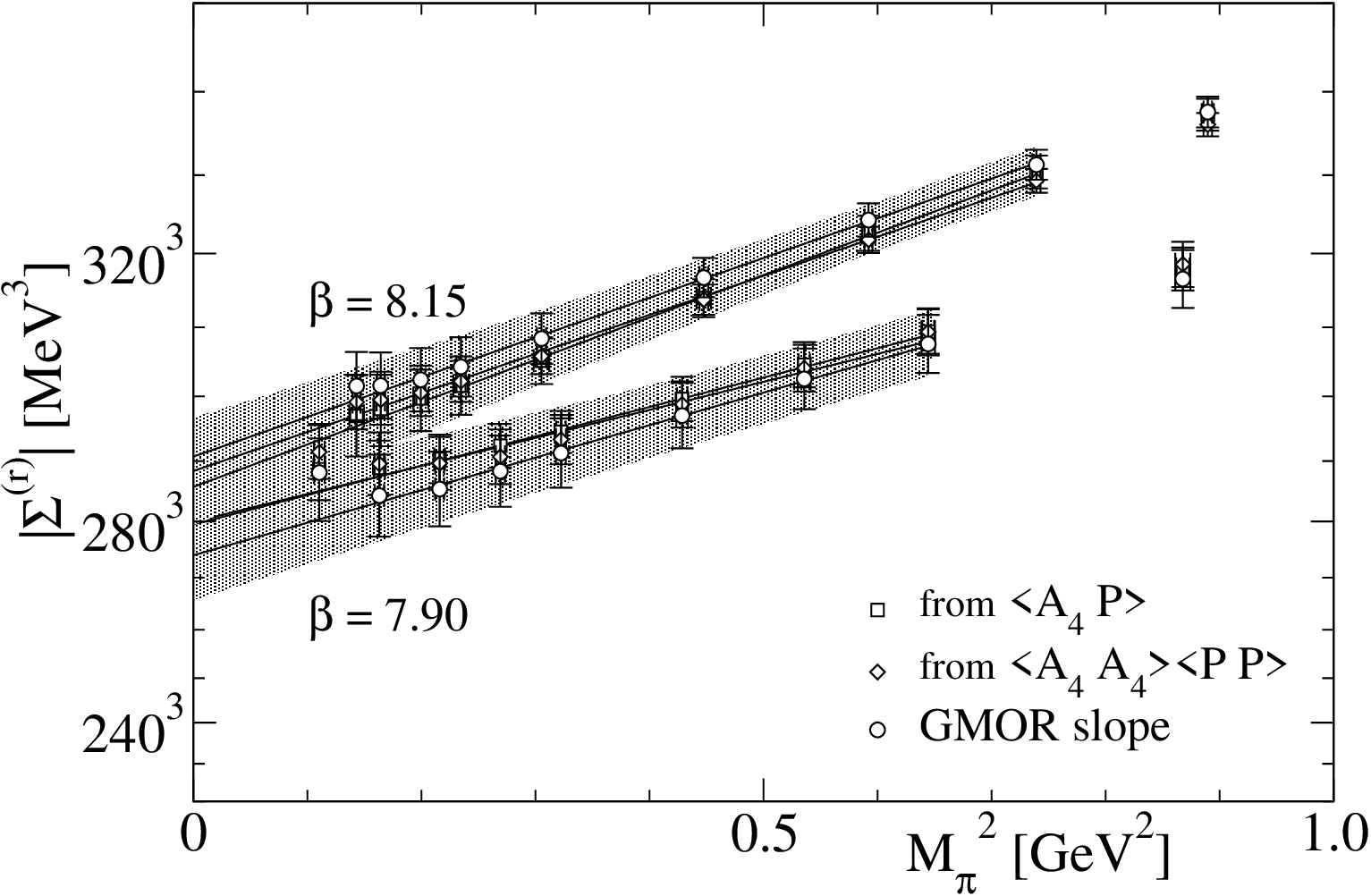}
\caption{\label{fig:sigma_vs_mpi2}
$|\Sigma^{(r)}|$ vs.\ $M_\pi^2$ as in Fig.\;\ref{fig:a3sigma_vs_am}, but in 
physical units.}
\end{figure}

Instead we have computed the renormalized condensate from the relations
(\ref{GMORrelation}), (\ref{eq:corrAP}) and (\ref{eq:corrAAPP}) which all contain
$\Sigma^{(r)}$. The first of these is the GMOR relation, the other two are
determinations directly from the coefficients  of propagators and implicitly
related to GMOR as well. In Figs.\ \ref{fig:a3sigma_vs_am} and
\ref{fig:sigma_vs_mpi2} we show our results for $\Sigma^{(r)}$ in lattice units as
a function of the bare quark mass. We find excellent agreement for all three
determinations. The values are consistent within the error bars. 

The dependence on the bare quark mass is compatible with the leading linear chiral
behavior.  Note, that we are not at small enough quark masses to be in the
so-called $\epsilon$-regime but are in the $p$-regime \cite{GaLe87a} (for
recent studies in the $\epsilon$-regime cf.\
\cite{HeJaLe99,GiHeLa0304,FuHaOk05}). We show the linear fit (with 1 s.d.\
error band) to the data from all three types of determination but omit the points
with smallest mass value in the fit (cf.\ our discussion of possible finite size 
effects).

\begin{figure}[tp]

\includegraphics*[width=8.3cm]{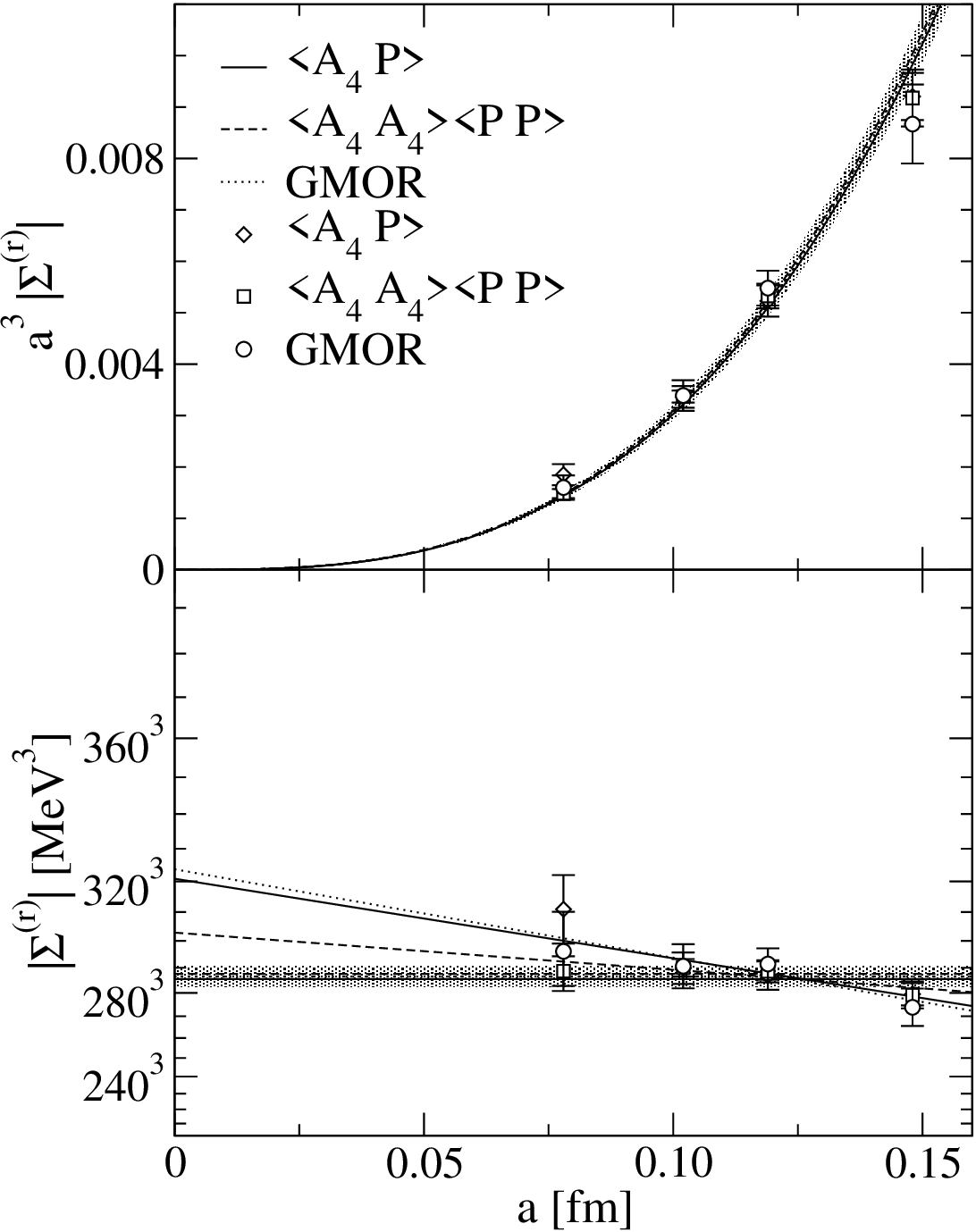}
\caption{\label{fig:2pack_sigma_vs_a}
$a^3\,|\Sigma^{(r)}|$ and $|\Sigma^{(r)}|$ vs.\ $a$. 
In the legends we indicate the quantities used in the derivation 
of the condensate value. In the bottom plot we show a constant extrapolation 
(horizontal error bands) and compare it to a linear one.
}
\end{figure}

So far we have restricted ourselves to the discussion of the largest physical
volume results for only two lattice spacings, corresponding to $\beta=7.90$ and
8.15. In order to analyze the scaling behavior in the continuum limit we show in
Fig.\;\ref{fig:2pack_sigma_vs_a} the values for all lattice spacings (always for the
largest volume) studied. Different physical lattice volumes are combined in this figure.
As may be seen from the upper part of the plot, the leading $a^3$-dependence is suggestive.

In the lower part of the figure we exhibit the renormalized condensate by dividing
out the expected leading $\mathcal{O}(a^3)$ scale dependence.
Since the Dirac action used obeys the GWC not exactly we cannot exclude linear corrections
to perfect scaling. In \cite{GaGoHa03a} these were found to be small for the hadron masses.
For the condensate some of these corrections are already taken into account by 
the (independently determined) $Z$-factors
in the determining equations.

For estimating the systematic error of the continuum extrapolation we therefore perform
both, a constant fit (no scaling corrections) and a linear fit to the
$a$-dependence. We do this for each type of derivation. The 
average of the resulting values for the constant fit is
$|\Sigma^{(r)}|=(286(4)~\textrm{MeV})^3$; the linear fit leads to a
larger value $|\Sigma^{(r)}|\approx(318(25)~\textrm{MeV})^3$.
For the final value we give the result of the constant fit but quote the difference
to the linear fit result as systematical error. Thus we find
\begin{equation} 
|\Sigma^{(r)}|=(286(4)(32)~\textrm{MeV})^3\;.
\end{equation} 
This value is slightly larger, although still
in the error limits of a determination from the overlap action \cite{BaBeGa05} and
larger than the corresponding results in Ref.\;\cite{ChHs03,BiSh05}. Our numbers are in
good agreement with calculations in the $\epsilon$-regime
\cite{GiHeLa0304,FuHaOk05}. They also agree with a recent continuum extrapolation
of the condensate based on the spectral distribution of the overlap
operator \cite{WeWi05}.

\subsection{Pion decay constant}

The pseudoscalar decay constants have been extracted from the asymptotic behavior
of the pseudoscalar correlation functions according to (\ref{eq:corrAA}) for pion
and kaon, respectively. (See also \cite{DaHeSp05} for another method to
determine $f_\pi$.)

\begin{figure}[tp]

\includegraphics*[width=8.3cm]{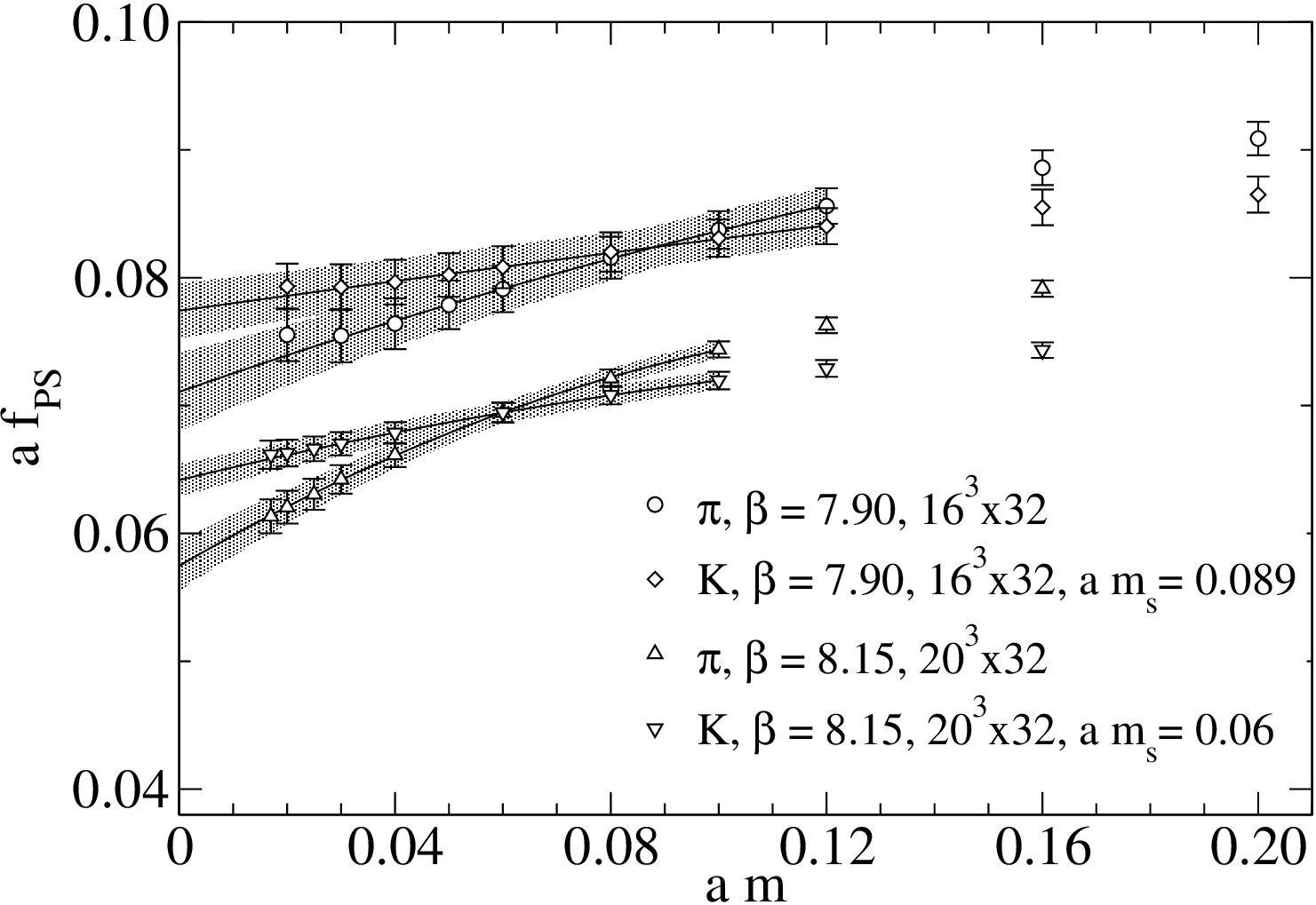}
\caption{\label{fig:afpiK_vs_am}
The dimensionless decay constants 
$a\,f_{\pi}$ and  $a\,f_K$, determined with Eq.\;(\ref{eq:corrAA}), vs.\ $a\,m$. 
}
~\\  
\includegraphics*[width=8.3cm]{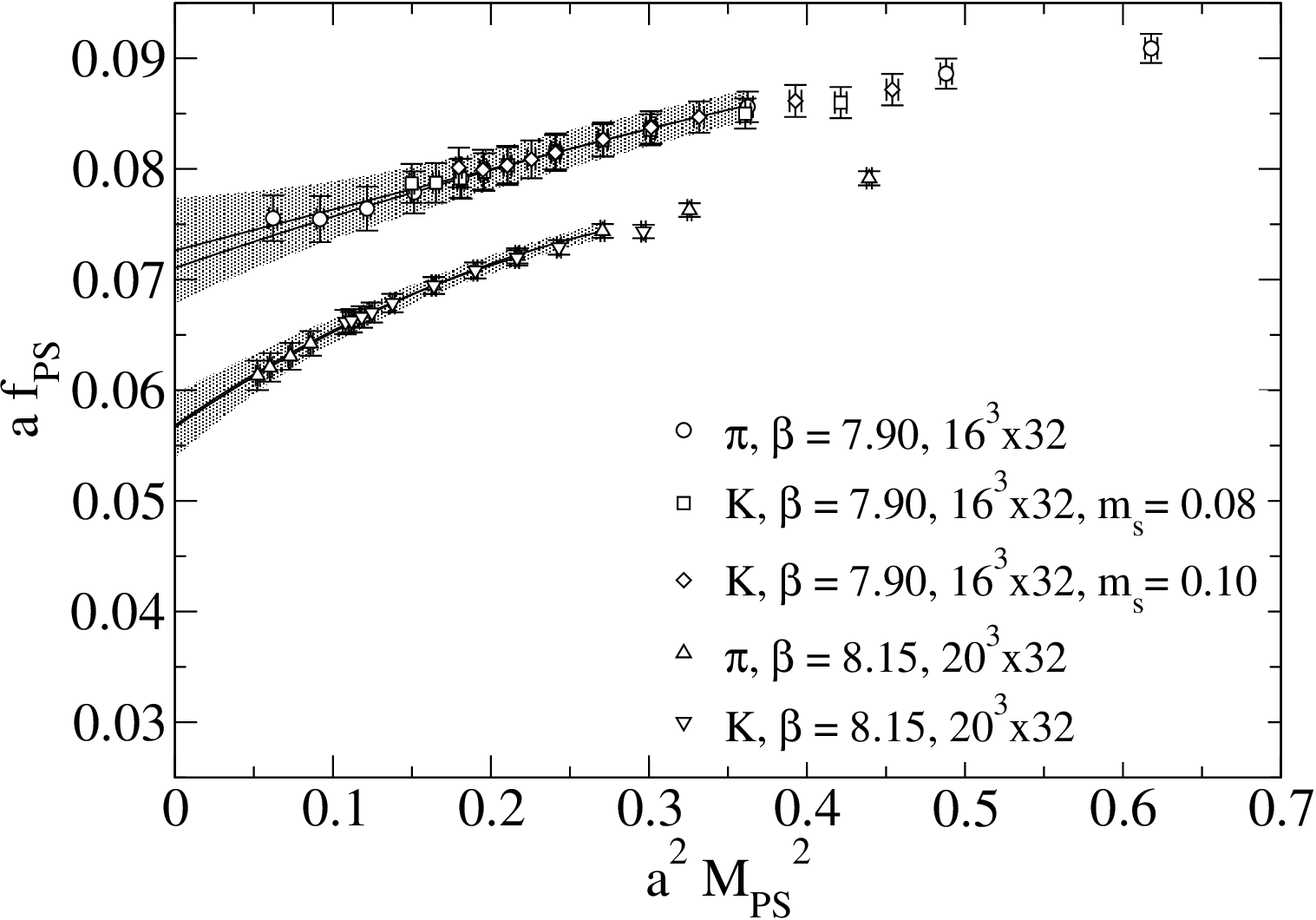}
\caption{\label{fig:afpiK_vs_a2mPS2}
$a\,f_{\pi,K}$ vs.\ $(a\,M_{PS})^2$ with error bands for chiral extrapolations as
discussed in the text. The fit includes only data in the indicated range.
}
\end{figure}

In Fig.\;\ref{fig:afpiK_vs_am} we compare these results. When plotting them as
functions of the respective pseudoscalar masses in Fig.\;\ref{fig:afpiK_vs_a2mPS2},
the data for pion and kaon essentially overlap each other and exhibit a universal
functional behavior. We also show the error band of a quadratic extrapolation to
the chiral limit.

For full QCD the chiral expansion of the pion decay constant should behave like
\cite{CoGaLe01}:
\begin{equation}\label{eq:chptfpiexpansion}
f_\pi/f    =  1+\xi\,\bar \ell_4+\mathcal{O}(\xi^2) \quad \textrm{with}\quad
\xi        =  \left( \frac{M_\pi}{4\,\pi\,F_\pi}\right)^2 \;.
\end{equation}
The value $\bar \ell_4=-\ln (M_\pi^2/ \Lambda^2)$ depends on the intrinsic QCD
scale $\Lambda$ and in \cite{CoGaLe01} it is suggested to use $\Lambda\approx
4\,\pi\,f_\pi$;  in Ref.\;\cite{AnCaCo04} a value of $\bar \ell_4\approx 4.0\pm
0.6$ is quoted.

ChPT also relates the decay constant to the scalar form factor radius via
\begin{equation}\label{eq:chptfpiscalrradius}
f_\pi/f=1 + \frac{1}{6}\langle r^2\rangle_s M_\pi^2 + \frac{13}{12}\xi+\mathcal{O}(\xi^2)\;.
\end{equation}
The relation holds for both, the pion and the kaon; this behavior is confirmed by our
results in Fig.\;\ref{fig:afpiK_vs_a2mPS2}.

Eq.\;(\ref{eq:chptfpiscalrradius}) may be translated to
\begin{equation}
\langle r^2\rangle_s=\frac{3}{8\,\pi^2\,f_\pi^2}\,
\left(\bar \ell_4 - \frac{13}{12}+\mathcal{O}(\xi)\right)\;.
\end{equation}
The authors of Ref.\ \cite{AnCaCo04} quote an expected value of $f_\pi/f=1.072(4)$
and $\langle r^2\rangle_s=0.61(4)\,\textrm{fm}^2$. 

In our quenched QCD case one expects correction terms with a logarithmic
singularity in the valence quark mass $m$. As pointed out in \cite{Sh92} the
leading order logarithmic term $ m\, \log m$ of ChPT involves quark loops that are
absent in the quenched case. There will be non-leading, e.g., logarithmic, terms, though. We therefore
allow in addition to the li\-near term in the quark mass $m$ also a term $m^2 \,\log
m$ in the extrapolating fit (cf.\ the discussion in \cite{DoDrHo04}). Actually, in
the fit it makes no significant difference whether we take this term or just $m^2$.

\begin{figure}[t]
\includegraphics*[width=8.3cm]{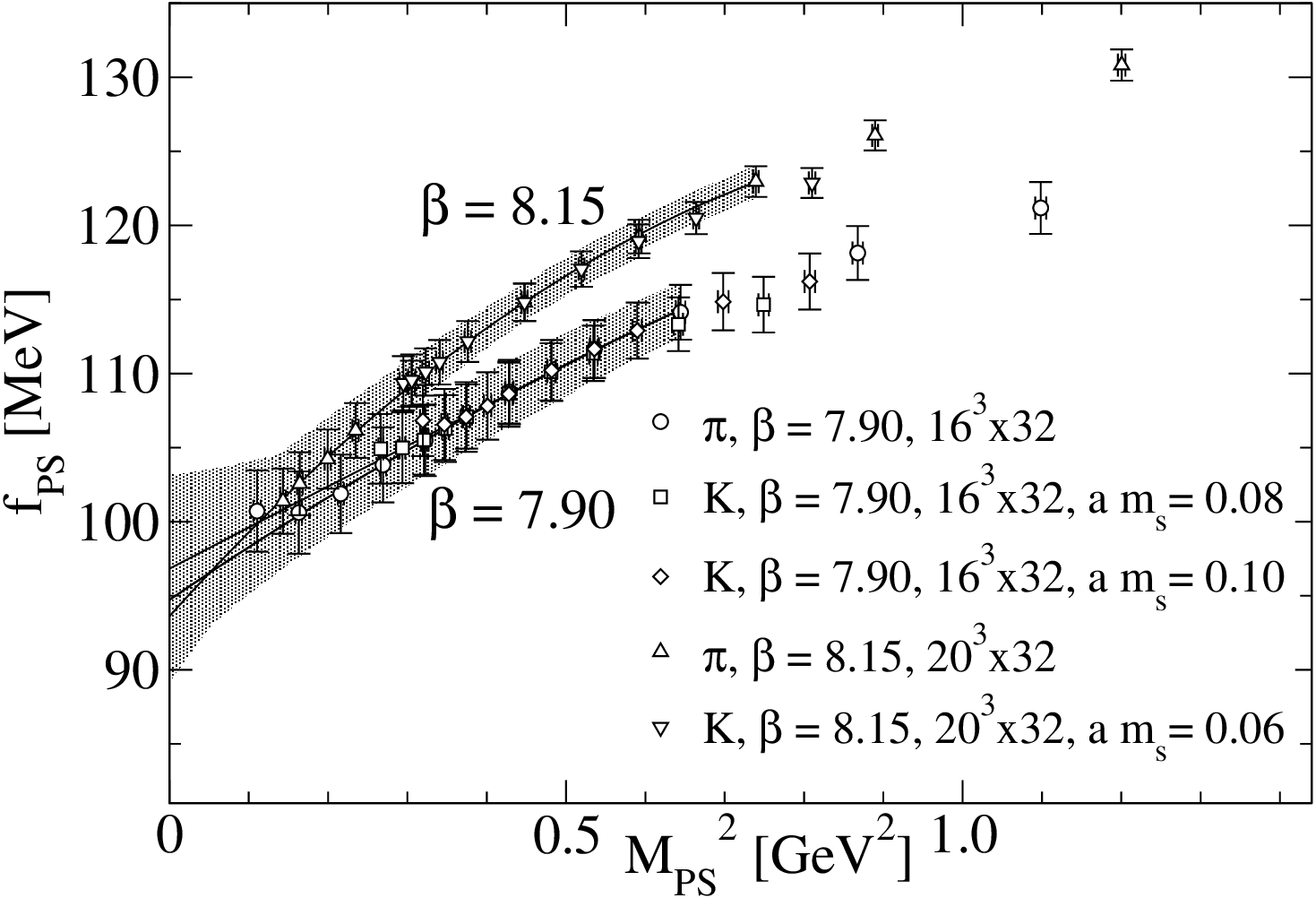}
\caption{\label{fig:fpiK_vs_mPS2}
Chiral limit of $f_{\pi,K}$ vs.\ $M_{PS}^2$. Error bands for quadratic extrapolation.}
\end{figure}

When removing the leading scale dependent factors, thus changing to physical
values, in Fig.\;\ref{fig:fpiK_vs_mPS2} we find still non-negligible 
$a$-dependence away from the chiral limit. Note that the physical values of $f_\pi$
and $f_K$ have to be read off at the respective different values of
$m_\textrm{\scriptsize PS}$.

Fig.\;\ref{fig:fpiK_vs_mPS2} (neglecting possible quenching effects) may be
analyzed according to the ChPT expansion (\ref{eq:chptfpiexpansion}). We see,
however, that the slope near the chiral limit shows considerable $a$-dependence.
Taking the slopes at face value we obtain values for $\langle r^2\rangle_s$ of 0.08
$\textrm{fm}^2$ and 0.13 $\textrm{fm}^2$, considerably smaller than expected for
full QCD.
 
\begin{figure}[tp]

\includegraphics*[width=8.3cm]{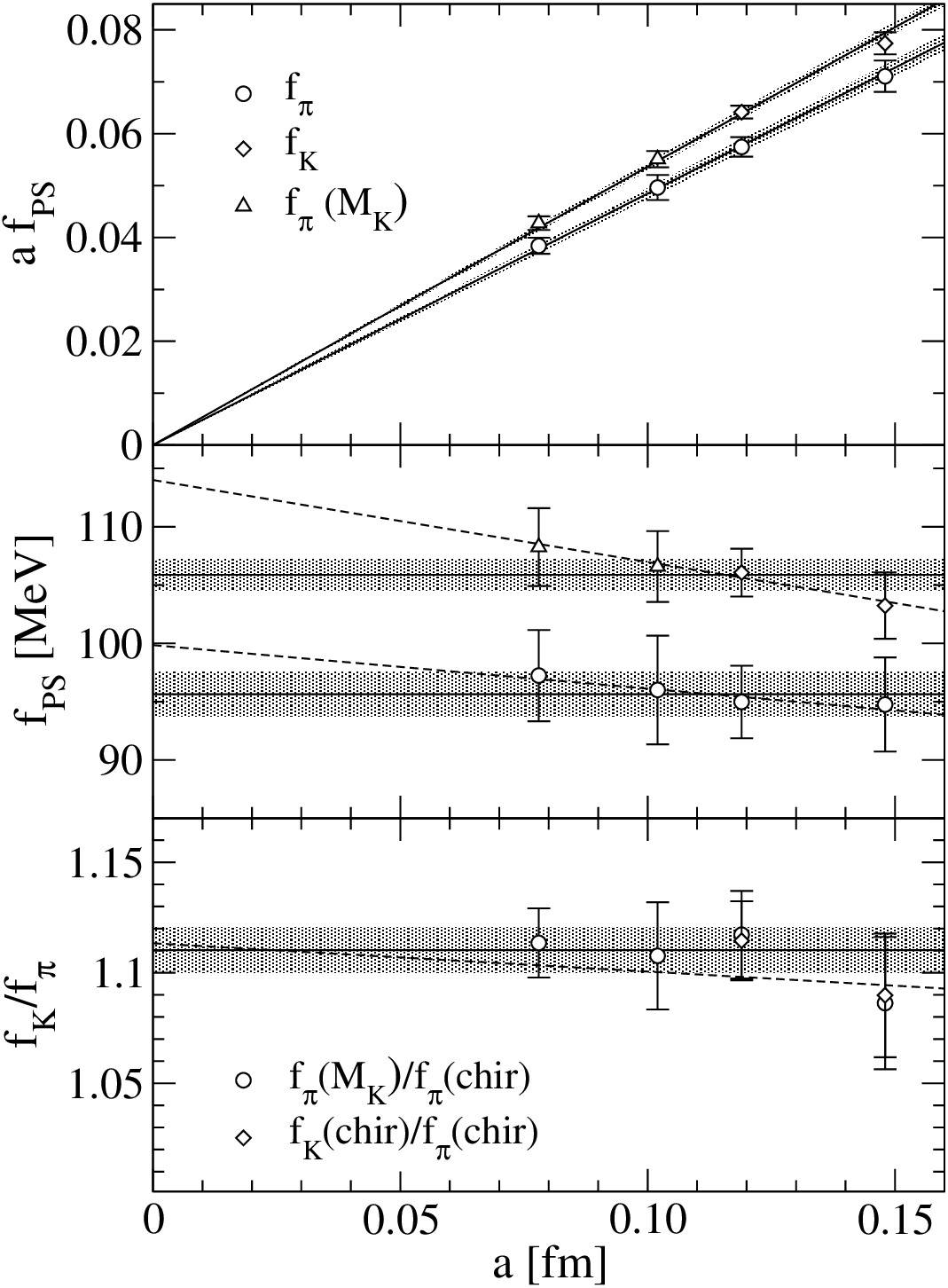}
\caption{\label{fig:3pack_fpi_vs_a}
$a\,f_{\pi,K}$, $f_{\pi,K}$ and $f_K/f_\pi$ vs.\ $a$. 
}
\end{figure}

For a better study of the scale dependence we plot in Fig.\;\ref{fig:3pack_fpi_vs_a} data
for all lattice spacings, always for the largest volumes studied here. This plot combines
different physical volumes.  As can be seen in the upper part of the figure, the expected
leading $a$-dependence is nicely exhibited. In the middle part of the figure we have removed
the leading $a$-dependence.

As mentioned above in our discussion of the condensate, we expect the linear  corrections
to the leading scaling behavior to be small, but cannot exclude them, since the action is
not an exact GW-operator. Like for the condensate we estimate the systematic error by
fitting the data to a constant (no scaling  violations) and a linear behavior in the
lattice spacing and quote the difference  as systematic error.  The constant extrapolation
for the pion gives $f_\pi=96(2)$ and the linear one gives $f_\pi=100(10)$. We therefore obtain as
final continuum extrapolation 
\begin{equation}
f_\pi=96(2)(4)\;\textrm{MeV}\;.
\end{equation}

For the kaon we have results for non-degenerate quark masses only at two lattice spacings (same
physical volume) and an extrapolation is thus underdetermined. At the other values of the
lattice spacing studied we therefore compute values for $f_{PS}$ from the $f_\pi$-results at
that mass $\overline m$, where at the pseudoscalar state for equal-mass quarks agrees with the
kaon mass. This is the usual method when only one quark mass is considered and is justified
from the universal behavior observed, e.g., in Fig.~\ref{fig:afpiK_vs_a2mPS2}. The
corresponding numbers are indicated by triangles in Fig.~\ref{fig:3pack_fpi_vs_a}.
Extrapolating in the same manner as discussed for the pion we end up with 
\begin{equation}
f_K=106(1)(8)\;\textrm{MeV}\;.
\end{equation}

We expect the leading corrections to scaling to cancel in the ratio $f_K/f_\pi$, plotted
in the lower part of Fig.~\ref{fig:3pack_fpi_vs_a}, and indeed the ratio is 
compatible with constant behavior. The continuum extrapolation along the lines discussed
above gives the result
\begin{equation}
f_K/f_\pi=1.11(1)(2)\;.
\end{equation}

Studies for the overlap action in quenched simulations have obtained
similar  results for $f_\pi$ \cite{DoDrHo04,GiHeLa0304,FuHaOk05,BiSh05,DuHo05,BaBeGa05}. The
experimental values are  $f_\pi= 92.4(0.3)\;\textrm{MeV}$ and 
$f_K=113.0(1.3)\;\textrm{MeV}$  \cite{PDBook04} (taking into account the factor
$\sqrt{2}$ in the definition).

\section{Conclusion}\label{sec:conclusion}

Within the Bern-Graz-Regensburg collaboration two Ginsparg-Wilson type Dirac
operators (FP fermions and the so-called chirally improved Dirac operator)  have
been studied, both obeying the GWC to a good approximation.  For the chirally
improved Dirac operator we know the quark bilinear renormalization constants
\cite{GaGoHu04}. This enables us to determine  basic low energy constants for the
light- and strange quark sector. All the results presented here are in the quenched
approximation, i.e., without taking into account dynamical sea quarks. Due to the
complicated and expensive form of the Ginsparg-Wilson type operators,  a full QCD
study still is a task for future work.

Since the chirally improved operator is substantially less expensive in
computational means than the overlap operator, it was possible to work at several
lattice spacings ranging from 0.15\;fm down to 0.08\;fm  and with different lattice
sizes. The results presented here are based mainly on simulations for lattice
spacing  0.15\;fm and 0.12\;fm on lattices with a physical spatial extent of
2.4\;fm. For an investigation of the scaling behavior and for studying finite size
effects we also consider results from the other lattice parameter sets with
different size and lattice constants. The data cover pion masses down to 330\;MeV.

In the range of our simulation parameters we cannot reliably identify behavior
specific for QChPT. Therefore the data have been extrapolated towards the chiral
limit with the leading order ChPT expansion for full QCD. All physical values have
been converted to the $\overline{\textrm{MS}}$-scheme (at scale 2 \textrm{GeV}).
We end up with the following final  (physical, renormalized) values:
\begin{equation}
\begin{array}{rccl}
\textrm{Quark masses:}   &\bar m  & = &4.1(2.4)\;\textrm{MeV}\;,\\
                         & m_s    & = &101(8)\;\textrm{MeV}\;,\\
\textrm{Condensate:}     &\Sigma & = &-(286(4)(32)~\textrm{MeV})^3\;,\\
\textrm{Decay constants:}&f_\pi   & = &96(2)(4)\;\textrm{MeV}\;,\\
                         &f_K     & = &106(1)(8)\;\textrm{MeV}\;,\\
                         &f_K/f_\pi& = &1.11(1)(2)\;.
\end{array}
\end{equation}
Like many quenched results, these numbers are surprisingly close to the
experimental values \cite{PDBook04}.

\begin{acknowledgments}
Support by Fonds zur F\"orderung der  Wissenschaftlichen Forschung in \"Osterreich
(P16824-N08 and P16310-N08) is gratefully acknowledged. The quark propagators have
been computed on the Hitachi SR8000 at the Leibniz  Rechenzentrum in Munich.  
\end{acknowledgments}

\clearpage

\appendix*
\section{Tables}

Here we collect our main results for the meson and  quark masses, for the 
condensate and the decay constants. All quantities are given as renormalized ones,
i.e., including the renormalization factors converting them to the
$\overline{\textrm{MS}}$-scheme at $\mu=2~\textrm{GeV}$.

\begin{table*}[htp]
\caption{\label{tab:meson_masses.pion.16x32_b7.90}%
The pion and kaon masses computed from Eq.\;\eqref{eq:corrPP} on the $16^3\times 32$, 
$\beta=7.90$ lattice. As soon as the light quark mass exceed the (fixed) strange quark
mass, the kaon becomes lighter than the pion.}
\begin{ruledtabular}
\begin{tabular}{l|llllll}
  $a\,m$ &    $a^2\,M_\pi^2$ &    $M_\pi^2[\textrm{GeV}^2]$ &    $M_\pi[\textrm{MeV}]$ &    $a^2\,M_K^2$ &    $M_K^2[\textrm{GeV}^2]$ &    $M_K[\textrm{MeV}]$ \\ \hline
  0.02 &      0.062(3) &	  0.110(5) &		 332(7)  &	    0.163(2) &        0.290(3) &   	   538(3) \\ 
  0.03 &      0.092(2) &	  0.163(4) &		 404(5)  &	    0.178(2) &        0.317(4) &   	   563(3) \\ 
  0.04 &      0.121(2) &	  0.216(4) &		 465(4)  &	    0.194(2) &        0.344(4) &   	   586(4) \\ 
  0.05 &      0.151(2) &	  0.269(4) &		 519(4)  &	    0.209(2) &        0.371(4) &   	   609(4) \\ 
  0.06 &      0.181(2) &	  0.322(4) &		 568(4)  &	    0.224(3) &        0.398(5) &   	   631(4) \\ 
  0.08 &      0.241(3) &	  0.429(5) &		 655(4)  &	    0.254(3) &        0.452(5) &   	   672(4) \\ 
  0.10 &      0.301(3) &	  0.536(6) &		 732(4)  &	    0.284(3) &        0.505(5) &   	   711(4) \\ 
  0.12 &      0.362(3) &	  0.644(6) &		 803(4)  &	    0.314(3) &        0.559(6) &   	   747(4) \\ 
  0.16 &      0.488(4) &	  0.868(7) &		 931(4)  &	    0.375(3) &        0.666(6) &   	   816(4) \\ 
  0.20 &      0.618(4) &	  1.098(7) &		 1048(3) &	    0.436(4) &        0.775(7) &   	   880(4) \\
\end{tabular}
\end{ruledtabular}
\caption{\label{tab:meson_masses.pion.20x32_b8.15}The pion and kaon masses computed
from Eq.\;\eqref{eq:corrPP} on the $20^3\times 32$, $\beta=8.15$ lattice.}
\begin{ruledtabular}
\begin{tabular}{l|llllll}
  $a\,m$ &    $a^2\,M_\pi^2$ &    $M_\pi^2[\textrm{GeV}^2]$ &    $M_\pi[\textrm{MeV}]$  &    $a^2\,M_K^2$ &    $M_K^2[\textrm{GeV}^2]$ &    $M_K[\textrm{MeV}]$ \\ \hline
  0.017 &	0.052(1) &	  0.143(4) &	    378(5)   & 0.108(2) &	 0.295(5) &	   543(4) \\ 
  0.02 &	0.060(1) &	  0.164(4) &	    405(4)   & 0.112(2) &	 0.305(5) &	   553(4) \\ 
  0.025 &	0.073(1) &	  0.199(4) &	    447(4)   & 0.118(2) &	 0.323(4) &	   568(4) \\ 
  0.03 &	0.086(1) &	  0.235(4) &	    484(4)   & 0.125(2) &	 0.341(4) &	   584(4) \\ 
  0.04 &	0.112(1) &	  0.305(4) &	    552(4)   & 0.138(1) &	 0.376(4) &	   613(3) \\ 
  0.06 &	0.164(1) &	  0.448(4) &	    669(3)   & 0.164(1) &	 0.448(4) &	   669(3) \\ 
  0.08 &	0.217(1) &	  0.592(4) &	    769(2)   & 0.190(1) &	 0.519(4) &	   721(3) \\ 
  0.10 &	0.271(1) &	  0.739(4) &	    860(2)   & 0.217(1) &	 0.592(4) &	   769(2) \\ 
  0.12 &	0.326(1) &	  0.890(4) &	    943(2)   & 0.243(1) &	 0.664(4) &	   815(2) \\ 
  0.16 &	0.439(2) &	  1.200(5) &	    1096(2)  & 0.297(1) &	 0.810(4) &	   900(2) \\
\end{tabular}
\end{ruledtabular}
\end{table*}

\begin{table*}[htp]
\caption{\label{tab:a3sigma.pion.16x32_b7.90} The light quark condensate as 
derived from different observables (lattice size $16^3\times 32$, $\beta=7.90$). In
the last line we give the extrapolation to the chiral limit as discussed in the
text.}
\begin{ruledtabular}
\begin{tabular}{lllllll}
 $a\,m$ & $a^3\,|\Sigma_{\langle A_4 P\rangle}|$ & $a^3\,|\Sigma_{\textrm{\scriptsize GMOR}}|$ & $a^3\,|\Sigma_{\langle P P\rangle\langle A_4 A_4\rangle}|$ & $\sqrt[3]{|\Sigma_{\langle A_4 P\rangle}|}[\textrm{MeV}]$ & $\sqrt[3]{|\Sigma_{\textrm{\scriptsize GMOR}}|}[\textrm{MeV}]$ & $\sqrt[3]{|\Sigma_{\langle P P\rangle\langle A_4 A_4\rangle}|}[\textrm{MeV}]$ \\ \hline
  0.02 &  0.0101(5) &  0.0101(8) & 0.0104(5) &        289(5) &        288(8) &        291(4) \\ 
  0.03 &  0.0102(5) &  0.0097(7) & 0.0102(5) &        289(4) &        284(7) &        290(5) \\ 
  0.04 &  0.0103(4) &  0.0098(6) & 0.0103(5) &        290(4) &        285(6) &        290(4) \\ 
  0.05 &  0.0105(4) &  0.0101(6) & 0.0104(5) &        292(3) &        288(6) &        291(4) \\ 
  0.06 &  0.0108(4) &  0.0104(6) & 0.0107(4) &        295(3) &        291(5) &        293(4) \\ 
  0.08 &  0.0113(4) &  0.0111(6) & 0.0113(4) &        300(3) &        297(5) &        299(3) \\ 
  0.10 &  0.0119(4) &  0.0117(5) & 0.0119(4) &        305(3) &        303(4) &        304(3) \\ 
  0.12 &  0.0125(4) &  0.0123(5) & 0.0125(4) &        309(3) &        308(4) &        309(3) \\ 
  0.16 &  0.0136(4) &  0.0134(5) & 0.0136(4) &        318(3) &        317(4) &        319(3) \\ 
  0.20 &  0.0146(4) &  0.0144(5) & 0.0146(4) &        326(3) &        324(4) &        326(3) \\ \hline
  chir. &            0.0092(5) &        0.0087(8) &      0.0092(6) &         279(5) &          274(8)   & 279(6)  \\

\end{tabular}
\end{ruledtabular}
~\\
~\\
\caption{\label{tab:sigma13.pion.20x32_b8.15}%
The light quark condensate as derived from different observables (lattice size
$20^3\times 32$, $\beta=8.15$).}
\begin{ruledtabular}
\begin{tabular}{lllllll}
 $a\,m$ & $a^3\,|\Sigma_{\langle A_4 P\rangle}|$ & $a^3\,|\Sigma_{\textrm{\scriptsize GMOR}}|$ & $a^3\,|\Sigma_{\langle P P\rangle\langle A_4 A_4\rangle}|$ & $\sqrt[3]{|\Sigma_{\langle A_4 P\rangle}|}[\textrm{MeV}]$ & $\sqrt[3]{|\Sigma_{\textrm{\scriptsize GMOR}}|}[\textrm{MeV}]$ & $\sqrt[3]{|\Sigma_{\langle P P\rangle\langle A_4 A_4\rangle}|}[\textrm{MeV}]$ \\ \hline
  0.017 &       0.0058(4) &	      0.0061(3) &	 0.0059(2)   &        297(6) &        302(5) &        299(4)\\ 
  0.02  &       0.0059(3) &	      0.0061(3) &	 0.0059(2)   &        298(6) &        302(5) &        299(4)\\ 
  0.025 &       0.0060(3) &	      0.0061(3) &	 0.0060(2)   &        300(5) &        302(5) &        301(3)\\ 
  0.03  &       0.0061(3) &	      0.0062(3) &	 0.0061(2)   &        302(4) &        304(4) &        302(3)\\ 
  0.04  &       0.0063(2) &	      0.0065(2) &	 0.0063(2)   &        305(3) &        308(4) &        306(3)\\ 
  0.06  &       0.0069(2) &	      0.0070(2) &	 0.0068(1)   &        314(3) &        317(3) &        314(2)\\ 
  0.08  &       0.0074(2) &	      0.0076(1) &	 0.0074(1)   &        323(2) &        324(2) &        322(2)\\ 
  0.10  &       0.0080(1) &	      0.0081(1) &	 0.0079(1)   &        330(2) &        331(2) &        329(1)\\ 
  0.12  &       0.0085(1) &	      0.0085(1) &	 0.0084(1)   &        337(2) &        338(2) &        336(1)\\ 
  0.16  &       0.0094(1) &	      0.0094(2) &	 0.0093(1)   &        349(2) &        349(2) &        348(1)\\ \hline
  chir. &          0.0052(3) &   0.0055(3) &	  0.0053(2) & 287(6) & 291(6) & 289(4) \\
\end{tabular}
\end{ruledtabular}
\end{table*}
\begin{table*}[htp]
\caption{\label{tab:decay_constants.pion.16x32_b7.90}Pion and kaon decay 
constants (from Eq.\;\eqref{eq:corrAA}) for $16^3\times 32$, 
$\beta=7.90$ lattice. In the last line we give the extrapolation to the 
(semi-)chiral limit (where the light quark masses vanish) as it is discussed in the text.}
\begin{ruledtabular}
\begin{tabular}{llllllll}
  $a\,m$ &     $a\,f_\pi/Z_A$ &    $a\,f_\pi$ &        $f_\pi[\textrm{MeV}]$ & $a\,f_K/Z_A$ &    $a\,f_K$ &        $f_K[\textrm{MeV}]$ \\ \hline
  0.02 &             0.076(2) &        0.076(2) &	   101(3) &	   0.080(2) &	     0.079(2) &        106(2) \\ 
  0.03 &             0.076(2) &        0.075(2) &	   101(3) &	   0.080(2) &	     0.079(2) &        106(2) \\ 
  0.04 &             0.077(2) &        0.076(2) &	   102(3) &	   0.080(2) &	     0.080(2) &        106(2) \\ 
  0.05 &             0.079(2) &        0.078(2) &	   104(3) &	   0.081(2) &	     0.080(2) &        107(2) \\ 
  0.06 &             0.080(2) &        0.079(2) &	   105(2) &	   0.082(2) &	     0.081(2) &        108(2) \\ 
  0.08 &             0.082(2) &        0.082(2) &	   109(2) &	   0.083(2) &	     0.082(2) &        109(2) \\ 
  0.10 &             0.084(1) &        0.084(1) &	   112(2) &	   0.084(1) &	     0.083(1) &        111(2) \\ 
  0.12 &             0.086(1) &        0.086(1) &	   114(2) &	   0.085(1) &	     0.084(1) &        112(2) \\ 
  0.16 &             0.089(1) &        0.089(1) &	   118(2) &	   0.086(1) &	     0.085(1) &        114(2) \\ 
  0.20 &             0.092(1) &        0.091(1) &	   121(2) &	   0.087(1) &	     0.087(1) &        115(2) \\ \hline
(semi-)chir. &       0.072(3) &       0.071(8) &        95(4)   &        0.078(2) &       0.077(2) &       103(3) \\
\end{tabular}
\end{ruledtabular}
\caption{\label{tab:decay_constants.pion.20x32_b8.15}Pion and kaon decay constants 
(from Eq.\;\eqref{eq:corrAA}) for $20^3\times 32$, 
$\beta=8.15$ lattice.}
\begin{ruledtabular}
\begin{tabular}{llllllll}
  $a\,m$ &     $a\,f_\pi/Z_A$ &    $a\,f_\pi$ &        $f_\pi[\textrm{MeV}]$ & $a\,f_K/Z_A$ &    $a\,f_K$ &        $f_K[\textrm{MeV}]$ \\ \hline
  0.017 &        0.062(1)  &	    0.061(1)  &        101(2) &       0.067(1)  &	 0.066(1)  &	    109(2) \\ 
  0.02 &         0.063(1)  &	    0.062(1)  &        103(2) &       0.067(1)  &	 0.066(1)  &	    110(2) \\ 
  0.025 &        0.064(1)  &	    0.063(1)  &        104(2) &       0.067(1)  &	 0.067(1)  &	    110(2) \\ 
  0.03 &         0.065(1)  &	    0.064(1)  &        106(2) &       0.0678(9) &	 0.0670(9) &	    111(1) \\ 
  0.04 &         0.067(1)  &	    0.0661(9) &        109(2) &       0.0686(8) &	 0.0679(8) &	    112(1) \\ 
  0.06 &         0.0703(8) &	    0.0695(8) &        115(1) &       0.0703(8) &	 0.0695(8) &	    115(1) \\ 
  0.08 &         0.0730(7) &	    0.0721(7) &        119(1) &       0.0716(7) &	 0.0708(7) &	    117(1) \\ 
  0.10 &         0.0752(6) &	    0.0744(6) &        123(1) &       0.0728(7) &	 0.0720(7) &	    119(1) \\ 
  0.12 &         0.0772(6) &	    0.0763(6) &        126(1) &       0.0737(7) &	 0.0729(7) &	    121(1) \\ 
  0.16 &         0.0801(6) &	    0.0792(6) &        131(1) &       0.0752(6) &	 0.0743(6) &	    123(1) \\ \hline
(semi-)chir. &  0.058(2)  &        0.058(2) &            95(3) &       0.065(1) &       0.064(1) &       106(2)  \\
\end{tabular}
\end{ruledtabular}
\end{table*}

\clearpage


\end{document}